\documentclass[12pt,aps,noshowpacs]{revtex4}
\usepackage{epsfig,amssymb,amsfonts,bm}
\usepackage{epsfig,amssymb,amsfonts,bm}
\usepackage[active]{srcltx}
\newcommand{\be}{\begin{equation}}
\newcommand{\ee}{\end{equation}}
\newcommand{\bea}{\begin{eqnarray}}
\newcommand{\eea}{\end{eqnarray}}
\newcommand{\beas}{\begin{eqnarray*}}
\newcommand{\eeas}{\end{eqnarray*}}
\newcommand{\ds}{\displaystyle}
\newcommand{\vep}{{\bm p}}

\newcommand{\veq}{{\bm q}}
\newcommand{\ves}{{\bm s}}

\newcommand{\veY}{{\bm Y}}
\newcommand{\ven}{{\bm n}}

\DeclareMathSymbol{\varGamma}{\mathord}{letters}{"00}
\makeatletter
\renewcommand{\@makefntext}[1]{\parindent=1em\noindent\hbox to 1.8em{\hss$^{\@thefnmark}$}#1}
\renewcommand{\@footnotemark}{\hbox{\mathsurround=0pt$^{\@thefnmark}$}}
\newcommand{\ftnote}[2]{\footnotemark[#1]\footnotetext[#1]{#2}}
\makeatother

\begin{document}

\title{Three-body $D\bar D\pi$ dynamics for the $X(3872)$}

\author{V. Baru}
\affiliation{Institut f\"ur Theoretische Physik II, Ruhr-Universit\"at Bochum, D-44780 Bochum, Germany
and Institute for Theoretical and Experimental Physics, B. Cheremushkinskaya 25, 117218 Moscow, Russia}

\author{C. Hanhart}
\affiliation{Forschungszentrum J\"ulich, Institute for Advanced Simulation, Institut f\"ur Kernphysik (Theorie) and J\"ulich Center for Hadron Physics, D-52425 J\"ulich, Germany}

\author{A. A. Filin}
\affiliation{Institut f\"ur Theoretische Physik II, Ruhr-Universit\"at Bochum, D-44780 Bochum, Germany
and Institute for Theoretical and Experimental Physics, B. Cheremushkinskaya 25, 117218 Moscow, Russia}

\author{Yu. S. Kalashnikova}
\affiliation{Institute for Theoretical and Experimental Physics, B. Cheremushkinskaya 25, 117218 Moscow, Russia}

\author{A. E. Kudryavtsev}
\affiliation{Institute for Theoretical and Experimental Physics, B. Cheremushkinskaya 25, 117218 Moscow, Russia}

\author{A. V. Nefediev}
\affiliation{Institute for Theoretical and Experimental Physics, B. Cheremushkinskaya 25, 117218 Moscow, Russia}

\begin{abstract}
We investigate the role played by the three-body $D\bar{D}\pi$ dynamics on the
near-threshold resonance $X(3872)$ charmonium state, which is assumed to be
formed by nonperturbative $D\bar D^*$ dynamics.
It is demonstrated that, as compared to the naive static-pions approximation,
the imaginary parts that originate from the inclusion of dynamical pions
reduce substantially the width from the $D\bar{D}\pi$ intermediate state. In particular, for a resonance peaked at
0.5~MeV below the $D^0\bar D^{*0}$ threshold, this contribution to the width is reduced by about a factor of 2,
and the effect of the pion dynamics on the width grows as long as the resonance
is shifted towards the $D^0\bar{D^0}\pi^0$ threshold.
Although the physical width of the $X$ is dominated by inelastic
channels, our finding should still be of importance for the $X$ line shapes
in the $D\bar{D}\pi$ channel below  $D{\bar D}^*$  threshold. For example, in the scattering length approximation,
the imaginary part of the scattering length 
includes effects of all the pion dynamics and does not only stem from the $D^*$ width. 
Meanwhile, we find that another important quantity for the $X$ phenomenology, the residue at the 
$X$ pole, is weakly sensitive to dynamical pions. In particular, we find that the binding energy 
dependence of this quantity from the full calculation is close to that found from a model with 
pointlike $D\bar D^*$ interactions only, consistent with earlier claims. Coupled-channel effects 
(inclusion of the charged $D\bar{D}^*$ channel) turn out to have a moderate impact on the results.

\end{abstract}

\maketitle

\section{Introduction}

Over the past decade we have witnessed fascinating progress in charmonium
spectroscopy, especially, due to the development of $B$-factories, the mass region
above the open-charm threshold became accessible for a systematic, high-statistics experimental
investigation. As a result, many new and unexpected states (the so-called
``$X, Y, Z$ states'') with unusual properties were discovered---for a recent
review see Ref.~\cite{HQWGreview}. Among these new
charmoniumlike states the $X(3872)$ meson found by Belle Collaboration in
2003 \cite{bellediscovery} is the best-studied state both experimentally and
theoretically. However, the $X(3872)$ still remains enigmatic, and there is no
consensus on the nature of this state.

The proximity of the $X$ to the $D^0 \bar D^{*0}$ threshold suggests the
dynamical (molecular) interpretation, though other options like $c \bar c$ or
tetraquark charmonium are discussed as well.
The $X(3872)$ was observed both in the $J/\psi \pi^+ \pi^-$ ($J/\psi \rho$) and $J/\psi \pi^+ \pi^- \pi^0$ ($J/\psi \omega$) modes
\cite{rho,omega}, which points to isospin violation in the wave function of the $X$. It is readily explained in the molecular picture as due to
the large (about $8$ MeV) mass difference between the charged and neutral $D
\bar D^*$ thresholds: the isospin violation is enhanced due to kinematical reasons, as the effective
phase space available in the case of the $\rho$ is much larger than that in the case
of the $\omega$ \cite{suzuki,oset}.
A molecular interpretation implies
the $1^{++}$ quantum numbers for the $X(3872)$ and, until recently, this
assignment was commonly accepted and supported by observation of the $X$ in
the $D^0 \bar D^{*0}$ decay mode
\cite{ddstarbelle1,ddstarbabar,ddstarbelle2}. However, while the analysis of
the $J/\psi \pi^+\pi^-$ decay mode of the $X(3872)$ yields either $1^{++}$ or
$2^{-+}$ quantum numbers \cite{rho}, the recent analysis of the $J/\psi
\pi^+\pi^-\pi^0$ mode seems to favour the $2^{-+}$ assignment \cite{omega},
though the $1^{++}$ option is not excluded. As shown in Ref.~\cite{1D2}, the
$X$ cannot be a naive $c \bar c$ $2^{-+}$ state and, were the $2^{-+}$ quantum
numbers confirmed, very exotic explanations for the $X$ would have to be
invoked. In the absence of such a confirmation we stick to the most
conventional $1^{++}$ assignment for the $X$.

Threshold affinity should lead to a significant admixture of the pertinent
charmed meson pair in the wave function of the resonance, whatever the nature
of the $X(3872)$ is, though it cannot {\it per se} shed any light on the
origin of binding mechanisms responsible for the formation of the $X$.
A natural explanation for the $X(3872)$ would be a $c \bar c$ $2^3P_1$
charmonium state ($\chi^{\prime}_{c1}$), residing at the $D^0 \bar D^{*0}$ threshold, but, unless
the coupling of the quark state to the charmed mesons channel is unnaturally
small, the naive bare $c \bar c$ spectrum should be distorted strongly by
coupled-channel effects. Indeed, the microscopic calculations
\cite{YuSK,simonov} confirm this pattern: the $X(3872)$ pole can be generated
dynamically by a strong coupling of the bare $\chi'_{c1}$ state to the $D \bar
D^*$ hadronic channel, with a large admixture of the $D \bar D^*$ component, see also
Ref.~\cite{Kang} where the fine-tuning of the $\chi'_{c1}$ state to the $D \bar
D^*$ threshold was discussed based on the analysis of line shapes for the $X(3872)$.

A competing approach is a traditional one-pion exchange (OPE) one. Historically,
long before the charmonium revolution of 2003, pion exchange between charmed
mesons was considered as a mechanism able to bind the isosinglet $D \bar D^{*}$
mesonic system and to form a deuteronlike state near threshold---see, for
example, Refs.~\cite{voloshin,tornqvist1}. Immediately after discovery of the
$X(3872)$, the OPE model was revisited
\cite{tornqvist2,swanson}. For the most recent work on the possibility for the
OPE to bind the $D \bar D^*$ system see Refs.~\cite{ch,ThCl}. Further
implications of the nearby pion threshold are discussed in
Refs.~\cite{braatenpions,pions}. In Refs.~\cite{ThCl,ch} divergent integrals
are made finite through the introduction of suitable form factors, and bound
states are found in the static approximation for the pion and neglecting the imaginary parts of the potential. Only the neutral
$D^0 \bar D^{*0}$ configuration was studied in Ref.~\cite{ch}, and the
dependence of the binding energy on the form factor cut off parameter
$\Lambda$ was investigated. It was shown that the bound state in the $D^0 \bar
D^{*0}$ system, with the binding energy around 1~MeV, exists only for the
values of $\Lambda$ of order of 6~GeV, that is for the values much larger than
admitted by interpretation of the form factors in terms of quark
models. The charged $D \bar D^*$ channel was included in Ref.~\cite{ThCl}, and it
was argued there that even for small cut offs of order of $1- 2$~GeV, a bound state with a binding energy of 1~MeV appears. This
result is interpreted then as a proof that the OPE provides
enough attraction to produce a bound state. Notice, however, that the $D^*D\pi$
coupling constant employed in calculations of Ref.~\cite{ThCl} is too large,
and is not compatible with the data on the $D^* D \pi$ decays.

The above-mentioned calculations treated the $D \bar D^*$ system in a
deuteronlike fashion: pions enter there in the form of a static
potential. There is, however, an important difference between the deuteron and
the $X$: the $D^{*0}$ mass is very close to the $D^0 \pi^0$ threshold. A
natural worry \cite{suzuki} is that in the $D \bar D^*$ system, bound by the
OPE, the pion may go on shell. The latter calls for the proper
inclusion of the three-body $D \bar D \pi$ unitarity cuts. As shown in
Ref.~\cite{deeply}, the cut effects are of paramount importance in the charmed
$D_{\alpha} \bar D_{\beta}$ system if one of the constituents has a large width,
dominated by the $S$-wave $D_{\beta} \to D_{\alpha} \pi$ decay: bound states
found in Ref.~\cite{closedeeply} in the static approximation disappear
completely from the spectrum if the full three-body treatment is invoked. In the
case of the $X(3872)$ 
the generic two-body $D\pi$ interaction goes via the $D^*$ and is in a $P$-wave. Due to this
``$P$-wave penalty'' one should not expect disastrous
consequences, though cut effects could distort strongly the resonance
shape. Indeed, inclusion of the $D^*$ finite width alone is known to
produce a spectacular bound-state peak in the $D^0 \bar{D}^0 \pi^0$ mass
distribution---see Refs.~\cite{lineshapeeric,stapleton,lineshapeours,ourX2}. This justifies a
full investigation of the role played by the three-body dynamics on a
near-threshold resonance, which is the subject of the present paper.

One of the  most important findings of our study is that the $X$-dynamics, and especially the
value of the effective coupling constant $X\to D\bar{D}^*$, is, in the molecular
scenario, completely fixed by the $X$ binding energy $E_B$, as long as
$E_B\ll \Delta M$, with $\Delta M=8.08$~MeV being the distance to the next
(charged) two-body threshold, in line with the properties of a true two-body
state, although there are various thresholds near by.
This is in line with the results of Ref.~\cite{Yasthreebody}, where
the interplay of scales was studied for the case of the presence of
$S$-wave interactions only.
At the same time, by an
explicit calculation, we have shown the validity of the central assumption
underlying the $X$-EFT~\cite{pions,XEFT}, namely, that pion effects can be
treated perturbatively for most observables. At leading order in such an EFT,
pions can be integrated out and predictions for the observables can be made
based on universal asymptotic behaviour of the $D\bar{D}^*$ wave function
\cite{Vol2004}.

The most striking effect of dynamical pions is seen in their effect on the
imaginary parts from the $D\bar{D}\pi$ intermediate states.
Specifically, the part of the $X$ width stemming from the width of the $D^*$ gets cut in half
once dynamical pions are included. A similar effect is observed once the three-body $D\bar D\pi$ cut
is accounted in the $D\bar D^*$ potential: the imaginary part of the $D\bar D^*$ potential gets reduced by a factor of more
than 2 compared to that for the static pion potential which leads to a further reduction of the $X$ width.
In total the width of the $X$ from the $D\bar{D}\pi$ intermediate states is reduced from 102 to 44 keV
due to the effects of dynamical pions. 
This observation could be of relevance for the $X$ line shapes below the elastic
threshold.

The paper is organised as follows. In Secs.~II A and B we introduce the notations and derive the system of
dynamical Faddeev-type equations for the $D\bar D^*$ scattering.
In Sec.~II C we apply these equations to the $X(3872)$ assuming $1^{++}$ assignment for this state.
Physical implications are discussed in Sec.~III. First, we compare the residues of the $D\bar D^*$ scattering amplitude
calculated with purely contact $D\bar D^*$ potential (with no pions) to our full dynamical results. We also
discuss the role of coupled-channel dynamics for the residue. Then, assuming that the $X(3872)$ has a resonance state
with the peak at $E_B=0.5$ MeV below the $D^0 \bar D^{*0}$ threshold, we calculate the $D^0 \bar D^0\pi^0$ line shape
within our fully dynamical treatment and compare it with various approximations.
The summary of the most important results is given in Sec.~IV.

\section{Three-body formalism}

\subsection{Kinematics and main definitions}

\begin{center}
\begin{figure}[t]
\centerline{\epsfig{file=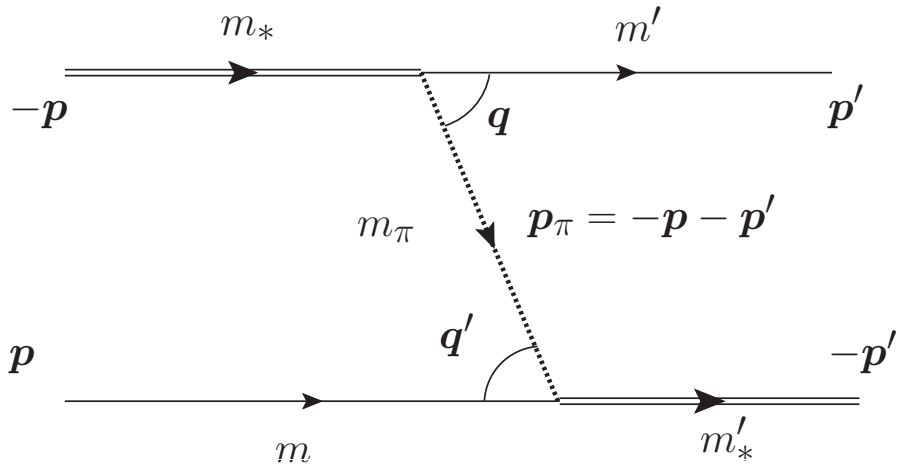, width=12cm}}
\caption{Kinematics of the $D \bar D^*$ scattering due to the OPE. Double lines denote $D^*$'s while single lines denote $D$'s.}\label{diagram}
\end{figure}
\end{center}

Consider first three channels:
\be
|2\rangle=D\bar{D}^*,\quad|\bar{2}\rangle=\bar{D}D^*,\quad|3\rangle=D\bar{D}\pi,
\label{chandef}
\ee
coupled by the OPE---see
Fig.~\ref{diagram} with $m'=m$ and $m_*'=m_*$ since, for a time being, we ignore isospins; the isospin structure of the
interaction will be considered in detail below. Here $m$ and $m_*$ are the mass of the 
$D(\bar{D})$ and the bare mass of the $D^*(\bar{D}^*)$, respectively.

The dynamical equations that emerge are basically a diagrammatic
representation
of those presented in Ref.~\cite{aay}.
In the centre-of-mass frame the momenta in the two-body systems $D \bar{D}^*$ and $\bar{D}D^*$ are defined as
\bea
\vep_D&=&\vep,\quad\vep_{\bar D^*}=-\vep,\\
\vep_{\bar D}&=&\bar {\vep},\quad\vep_{D^*}=-\bar {\vep},
\eea
while, in the three-body $D\bar{D}\pi$ system, the momenta can be defined in terms of two sets of Jacobi variables, $\{\vep,\veq\}$ and $\{\bar{\vep},\bar{\veq}\}$:
\be
\vep_D=\vep,\quad\vep_{\bar D}=-\veq-\frac{m}{m+m_{\pi}}\vep,\quad\vep_{\pi}=\veq-\frac{m_{\pi}}{m+m_{\pi}}\vep,
\ee
or
\be
\vep_D=-\bar{\veq}-\frac{m}{m+m_{\pi}}\bar{\vep},\quad
\vep_{\bar D}=\bar {\vep},\quad\vep_{\pi}=\bar{\veq}-\frac{m_{\pi}}{m+m_{\pi}}\bar{\vep}.
\ee
Jacobi variables belonging to different sets are related to
each other as
\bea
\bar{\veq}=\alpha \veq+\beta \vep,\quad\bar {\vep}=-\veq-\alpha \vep,\\
\veq=\alpha \bar {\veq}+\beta \bar {\vep},\quad\vep=-\bar {\veq}-\alpha\bar{\vep},
\eea
where
$$
\alpha=\frac{m}{m+m_{\pi}},\quad\beta=\alpha^2-1=-\frac{(2m+m_{\pi})m_{\pi}}{(m+m_{\pi})^2}.
$$

The $D\bar{D}^*\pi$ vertex is
\be
v_{D\bar{D}^*\pi}(\veq)=g\; {\bm \epsilon}\cdot \veq,
\label{g1}
\ee
where ${\bm \epsilon}$ is the $D^*$ polarisation vector, $\veq$ is the relative
momentum in the $D\pi$ system, and $g$ is the coupling constant---in
general here one could also introduce a form factor, however, since we are interested in near-threshold phenomena,
such a form factor would not change the general properties of the results.
The constant $g$ can be fixed from the $D^{*0}\to D^0\pi^0$ width via
\be
\varGamma(D^{*0}\to D^0\pi^0)\equiv\varGamma_*=\frac{8\pi^2}{3}g^2\mu_q(D^0\pi^0)\left[2\mu_q(D^0\pi^0)(m_{0*}-m_0-m_{\pi^0})\right]^{3/2},
\label{D*width}
\ee
where the reduced mass is defined as
\be
\mu_q(XY)=\frac{m_Xm_Y}{m_X+m_Y}.
\ee

Then the channels (\ref{chandef}) communicate via the interaction potentials
\bea
V_{32}^{m}(\vep,\veq;\vep')&=&g q_m\delta(\vep-\vep'),\label{V32}\\
V_{3 \bar 2}^{m}(\bar {\vep},\bar {\veq};\bar {\vep}')&=&g\bar{q}_m\delta(\bar {\vep}-\bar {\vep}'),\label{V3a2}
\eea
and similarly for $V_{23}^{m}$ and $V_{\bar 2 3}^{m}$.

The inverse two- and three-body free propagators are defined by
\be
D_2(\vep)=m+m_*+\frac{p^2}{2\mu_*}-M,\quad D_3(\vep,\veq)=2m+m_{\pi}+\frac{p^2}{2\mu_p}+\frac{q^2}{2\mu_q}-M,
\label{g0}
\ee
with $M$ being the mass of the system, which can be expressed, for example, in terms of the energy related to the
$D\bar{D}^*$ threshold, while the reduced masses are defined as
\be
\mu_*=\frac{mm_*}{m+m_*},\quad\mu_p=\frac{m(m+m_{\pi})}{2m+m_{\pi}},
\quad\mu_q \equiv \mu_q(D\pi)=\frac{mm_{\pi}}{m+m_{\pi}},
\ee
and, nonrelativistically, $\mu_*=\mu_p$.

Finally, we define the bare $D\pi$ self-energy $\Sigma(p)$ as
\be
\Sigma(p)=\frac{g^2}{3}\int \frac{q^2d^3q}{D_3(\vep,\veq)}.
\label{G2def}
\ee

\subsection{Three--body equation for the $t$ matrix: One-pion exchange potential}

The dynamical equation for the $t$ matrix $t_{ik}^{mn}(\vep,\vep')$, where $i,k=2,\bar{2},3$ and $m$ and $n$ are the Lorentz indices, reads schematically:
\be
t=V-V{\cal G}_0t,
\label{fulleq}
\ee
where ${\cal G}_0$ is the diagonal matrix of the free Green's functions (\ref{g0}) and the interaction potential $V$ possesses just two nontrivial components given by Eqs.~(\ref{V32}) and (\ref{V3a2}). The system of nine equations (\ref{fulleq}) splits into three decoupled subsystems:
\be
\left\{
\begin{array}{rcl}
t_{22}&=&\ds -V_{23}\frac1{D_3}t_{32}\\
t_{\bar 2 2}&=&\ds-V_{\bar 2 3}\frac1{D_3}t_{32}\\
t_{32}&=&\ds V_{32}-V_{32}\frac1{D_2}t_{22}-V_{3 \bar 2}\frac1{D_2}t_{\bar 2 2},
\end{array}
\right.
\label{s10}
\ee
\be
\left\{
\begin{array}{rcl}
t_{2 \bar 2}&=&\ds-V_{23}\frac1{D_3}t_{3 \bar 2}\\
t_{\bar 2 \bar 2}&=&\ds-V_{\bar 2 3}\frac1{D_3}t_{3 \bar 2}\\
t_{3 \bar 2}&=&\ds V_{3 \bar 2}-V_{32}\frac1{D_2}t_{2 \bar 2}-V_{3 \bar 2}\frac1{D_2}t_{\bar 2\bar 2},
\end{array}
\right.
\label{s20}
\ee
\be
\left\{
\begin{array}{rcl}
t_{23}&=&\ds V_{23}-V_{23}\frac1{D_3}t_{33}\\
t_{\bar 2 3}&=&\ds V_{\bar 2 3}-V_{\bar 2 3}\frac1{D_3}t_{33}\\
t_{33}&=&\ds-V_{32}\frac1{D_2}t_{23}-V_{3 \bar 2}\frac1{D_2}t_{\bar 23}.
\end{array}
\right.
\label{s30}
\ee

Below, in this chapter, we give schematically the derivation of the
three-body equations for the $t$ matrix which follows from
Eqs.~(\ref{s10})-(\ref{s30}). To simplify notations distinguish neither
between $V_{\bar{2}3}$ and $V_{23}$ nor between $V_{3\bar{2}}$ and $V_{32}$.
We also omit all arguments. The full and detailed derivation can be found in the Appendix.

In the systems (\ref{s10}) and (\ref{s20}) we exclude the third equation and, using the definition (\ref{G2def}), arrive at
\be
\left\{
\begin{array}{rcl}
t_{22}&=&\ds-\Sigma+\Sigma\frac1{D_2}t_{22}+V_{23}\frac1{D_3}V_{32}\frac1{D_2}t_{\bar{2}2}\\[3mm]
t_{\bar{2}2}&=&\ds-V_{\bar{2}3}\frac1{D_3}V_{32}+\Sigma\frac1{D_2}t_{\bar{2}2}+V_{\bar{2}3}\frac1{D_3}V_{32}\frac1{D_2}t_{22},
\end{array}
\right.
\label{s102}
\ee
and a similar system for the components $t_{2\bar{2}}$ and $t_{\bar{2}\bar{2}}$, which we do not quote here.
Note that the relations between the breakup amplitudes $t_{23}$ and $t_{\bar{2}3}$ and the two-body 
amplitudes $t_{2\bar{2}}$ and $t_{\bar{2}\bar{2}}$ are given in Appendix A. 

The interaction respects $C$ parity, so it is possible to define $C$-even and
$C$-odd $D \bar D^*$ matrix elements. Indeed, the system (\ref{s102}) can be
rewritten for the combinations
\be
t_{\pm}=t_{22}\pm t_{\bar{2}2},
\ee
which satisfy then the following equations:
\be
\Delta \frac1{D_2} t_\pm=-\Sigma\mp V_{23}\frac1{D_3}V_{32}\pm V_{23}\frac1{D_3}V_{32}\frac1{D_2}t_\pm,
\ee
where the inverse dressed $D^*$ propagator $\Delta(p)$ is introduced as
\be
\Delta(p)=m_*+m+\frac{p^2}{2\mu_*}-M-\Sigma(p).
\label{Deltadef}
\ee

Finally, substituting
\be
t_{\pm}=-\frac{\Sigma D_2}{\Delta}+\frac{D_2}{\Delta}a_{\pm}\frac{D_2}{\Delta},
\label{intra0}
\ee
one arrives at the following equation for the new function $a_{\pm}^{mn}(\vep,\vep')$:
\be
a_{\pm}=V_{\pm}-V_{\pm}\Delta^{-1}a_{\pm},
\label{a0}
\ee
which, in the full form (see the Appendix), reads
\be
a_{\pm}^{mn}(\vep,\vep',E)=V^{mn}_{\pm}(\vep,\vep')
-\int d^3s V^{mp}_{\pm}(\vep,\ves) \Delta^{-1}(s)a_{\pm}^{pn}(\ves,\vep',E),
\label{a}
\ee
where the generic OPE potential has the form:
\be
V^{mn}_{\pm}(\vep,\vep')=\mp g^2\frac{(\vep'+\alpha\vep)_m (\vep+\alpha\vep')_n}{D_3(\vep,\vep')},
\quad \alpha=\frac{m}{m_{\pi}+m}.
\label{Vmn}
\ee

In what follows we confine ourselves to the $C$-even states only and therefore we consider only the amplitude $a_+^{mn}(\vep,\vep')\equiv a^{mn}(\vep,\vep')$ and the corresponding potential $V_+^{mn}(\vep,\vep')\equiv V^{mn}(\vep,\vep')$.

Equation~(\ref{a}) was derived neglecting
isospin. However, in the case of the $X(3872)$, the two two-body thresholds, the
neutral $D^0\bar{D}^{*0}+\mbox{c.c.}$ threshold and the charged one
$D^+ D^{*-}+\mbox{c.c.}$, are split by only around 8~MeV, so they are both
potentially relevant, and Eq.~(\ref{a}) is to be modified accordingly.

For antimesons, we stick to the convention of Ref.~\cite{ThCl}, that is we define
\be
|\bar M\rangle={\hat C} |M\rangle,
\ee
where $\hat{C}$ is the $C$-parity transformation operator, and $|M\rangle$ ($|\bar{M}\rangle$) denotes the wave function of the meson (antimeson).
Then interpolating currents of the pions and the $D$ and $D^*$ meson involved are chosen as
\bea
&&\pi^0=\frac{1}{\sqrt{2}}(\bar{u}\gamma_5 u-\bar{d}\gamma_5 d),\quad\pi^+=\bar{d}\gamma_5 u,\quad\pi^-=\bar{u}\gamma_5d,\\
&&D^0=\bar{u}\gamma_5 c,\quad \bar{D}^0=\bar{c}\gamma_5 u,\quad D^+=\bar{d}\gamma_5 c,\quad D^-=\bar{c}\gamma_5 d,\\
&&D^{*0}=\bar{u}\gamma_\mu c,\quad\bar{D}^{*0}=-\bar{c}\gamma_\mu u,\quad D^{*+}=\bar{d}\gamma_\mu c,\quad
D^{*-}=-\bar{c}\gamma_\mu d.
\label{dstar}
\eea

Since the pion is an isovector, the isospin structure of the OPE
potential (\ref{Vmn}) is given by the product $\vec\tau_1\cdot\vec\tau_2$.
Therefore, in order to evaluate the isospin coefficients, we are to compute the matrix element $\langle i|\vec\tau_1\cdot\vec\tau_2|k\rangle$ for $i,k=0$, $\bar{0}$, $c$, and $\bar{c}$, where the latter states are defined as:
\be
|0\rangle=D^0\bar{D}^{*0},\quad|\bar{0}\rangle=\bar{D}^0 D^{*0},\quad|c\rangle=D^+ D^{*-},\quad|\bar{c}\rangle=
D^- D^{*+}.
\ee
This will define the full OPE potentials $V^{mn}_{ik}(\vep,\vep')$.
Notice that, since strong interactions respect $C$-parity, then the following relation holds:
\be
\langle i|\vec\tau_1\cdot\vec\tau_2|\bar{k}\rangle=\langle \bar{i}|\vec\tau_1\cdot\vec\tau_2|k\rangle,\quad i,k=0,c.
\label{CV}
\ee

It is easy to find then for the two nonvanishing coefficients:
\begin{eqnarray}
&&\langle 0|\vec\tau_1\cdot\vec\tau_2|\bar{0}\rangle=\langle c|\vec\tau_1\cdot\vec\tau_2|\bar{c}\rangle=1\quad\mbox{(neutral pion exchange)},\nonumber\\[-4mm]
\label{isocoef}\\[-4mm]
&&\langle 0|\vec\tau_1\cdot\vec\tau_2|\bar{c}\rangle=\langle c|\vec\tau_1\cdot\vec\tau_2|\bar{0}\rangle=2\quad\mbox{(charged pion exchange)}.\nonumber
\end{eqnarray}

It is clear therefore that nonvanishing OPE potentials are (see Fig.~\ref{diagram}):
\begin{eqnarray}
V^{mn}_{0 \bar 0}(\vep,\vep')&=&V^{mn}_{\bar 0 0}(\vep,\vep'),~\mbox{with}~m=m'=m_0,~m_{\pi}=m_{\pi^0},\label{V1}\\
V^{mn}_{c \bar c}(\vep,\vep')&=&V^{mn}_{\bar c c}(\vep,\vep'),~\mbox{with}~m=m'=m_c,~m_{\pi}=m_{\pi^0},\\
V^{mn}_{c \bar 0}(\vep,\vep')&=&V^{mn}_{\bar c 0}(\vep,\vep'),~\mbox{with}~m=m_0,~m'=m_c,~m_{\pi}=m_{\pi^c},\\
V^{mn}_{0 \bar c}(\vep,\vep')&=&V^{mn}_{\bar 0 c}(\vep,\vep'),~\mbox{with}~m=m_c,~m'=m_0,~m_{\pi}=m_{\pi^c},\label{V4}
\end{eqnarray}
where $m_0$, $m_c$, $m_{\pi^0}$, and $m_{\pi^c}$ are the masses of the neutral
and charged $D$ meson, and the pions, respectively. The factor of 2 needed for
charge-exchange potentials [see Eq.~(\ref{isocoef})] will be introduced in the
scattering equations explicitly.

The explicit form of the OPE potentials can be obtained then as the generalisation of the generic potential (\ref{Vmn}):
\be
V^{mn}_{ik}(\vep,\vep')=(\vep'+\alpha_{ik} \vep)^m(\vep+\alpha'_{ik}\vep')^nF_{ik}(\vep,\vep'),
\quad F_{ik}(\vep,\vep')=-\frac{g^2}{D_{3ik}(\vep,\vep')},
\label{Vmn2}
\ee
with
\beas
\alpha_{00}=\alpha'_{00}=\frac{m_0}{m_{\pi^0}+m_0},\quad \alpha_{cc}=\alpha'_{cc}=\frac{m_c}{m_{\pi^0}+m_c},\\
\alpha_{0c}=\alpha'_{c0}=\frac{m_c}{m_{\pi^c}+m_c},\quad\alpha_{c0}=\alpha'_{0c}=\frac{m_0}{m_{\pi^c}+m_0},
\eeas
and
\begin{eqnarray}
D_{300}(\vep,\vep')&=&2m_0+m_{\pi^0}+\frac{p^2}{2m_0}+\frac{p'^2}{2m_0}+\frac{(\vep+\vep')^2}
{2m_{\pi^0}}-M-i0,\nonumber\\
D_{3cc}(\vep,\vep')&=&2m_c+m_{\pi^0}+\frac{p^2}{2m_c}+\frac{p'^2}{2m_c}+\frac{(\vep+\vep')^2}
{2m_{\pi^0}}-M-i0,\label{D3s}\\
D_{30c}(\vep,\vep')&=&m_c+m_0+m_{\pi^c}+\frac{p^2}{2m_0}+\frac{p'^2}{2m_c}+\frac{(\vep+\vep')^2}
{2m_{\pi^c}}-M-i0,\nonumber\\
D_{3c0}(\vep,\vep')&=&m_c+m_0+m_{\pi^c}+\frac{p^2}{2m_c}+\frac{p'^2}{2m_0}+\frac{(\vep+\vep')^2}
{2m_{\pi^c}}-M-i0.\nonumber
\end{eqnarray}

The 16-component OPE interaction potential $V^{mn}_{ik}(\vep,\vep')$ gives rise to a system of coupled equations for the 16 components of the $D\bar{D}^*$ scattering $t$ matrix. However, due to symmetries of the OPE potential (\ref{V1})--(\ref{V4}), many $t$ matrix components coincide with one another. In particular, it is easy to demonstrate that \be
a^{mn}_{ik}=a^{mn}_{\bar i \bar k},\quad a^{mn}_{\bar i k}=a^{mn}_{i \bar k},\quad i,k=0,c,
\ee
so that only eight independent $t$ matrix components remain and split into two groups, four components in each, which satisfy two disentangled subsystems of equations. In what follows we shall be interested in the $|0\rangle$ final state, so we consider only the first subsystem of equations of these two, which reads
\be
\left\{
\begin{array}{rcl}
a_{00}^{mn}(\vep,\vep',E)=V_{00}^{mn}(\vep,\vep')&-&\ds
\int\frac{d^3s}{\Delta_0(s)}V_{00}^{mp}(\vep,\ves)a_{00}^{pn}(\ves,\vep',E)\\[2mm]
&-&\ds 2\int\frac{d^3s}{\Delta_c(s)}V_{0c}^{mp}(\vep,\ves)a_{c0}^{pn}(\ves,\vep',E)\\[3mm]
a_{c0}^{mn}(\vep,\vep',E)=2V_{c0}^{mn}(\vep,\vep')
&-&\ds 2\int\frac{d^3s}{\Delta_0(s)}V_{c0}^{mp}(\vep,\ves)a_{00}^{pn}(\ves,\vep',E)\\[2mm]
&-&\ds \int\frac{d^3s}{\Delta_c(s)}V_{cc}^{mp}(\vep,\ves)a_{c0}^{pn}(\ves,\vep',E).
\end{array}
\right.
\label{aa}
\ee

\subsection{$1^{++}$ channel}\label{fullsec}

From now on we stick to the quantum numbers inherent to the $X(3872)$
charmonium, assuming the latter to be $1^{++}$ as per discussion in the Introduction.
To perform the partial-wave decomposition of the scattering
equation (\ref{aa}) we expand the amplitude and the potential in terms of the
spherical vectors $\veY_{JLM}(\ven)$: \be
a_{ik}^{mn}(\vep,\vep',E)=\sum_J\sum_{L_1L_2}a_{ik}^{J,L_1,L_2}(p,p',E)\sum_M(\veY_{JL_1M}(\ven))^m(\veY^*_{JL_2M}(\ven'))^n,
\ee \be
V_{ik}^{mn}(\vep,\vep')=\sum_J\sum_{L_1L_2}V_{ik}^{J,L_1,L_2}(p,p')\sum_M(\veY_{JL_1M}(\ven))^m(\veY^*_{JL_2M}(\ven'))^n,
\ee where $\ven$ and $\ven'$ are the unit vectors for the momenta $\vep$ and
$\vep'$, respectively, and, given the chosen quantum numbers, the relevant
matrix elements are those with $J=1$ and $L_1,L_2=0,2$. To simplify notations,
we omit everywhere the superscript for the total momentum $J=1$.

Furthermore, once we are interested only in the $S$-wave in final state, we need to know only the $a_{ik}^{SS}$ and $a_{ik}^{DS}$ matrix elements of the amplitude, so that we use the following decomposition of the amplitude:
\be
a_{ik}^{mn}(\vep,\vep',E)=a_{ik}^{SS}(p,p',E)T_{SS}^{mn}+a_{ik}^{DS}(p,p',E)T_{DS}^{mn},
\ee
where, using properties of the spherical vectors, one can find for the projectors:
\be
T_{SS}^{mn}=\frac{1}{4\pi}\delta_{mn},\quad T_{DS}^{mn}=\frac{1}{4\pi\sqrt{2}}
(\delta_{mn}-3n_mn_m)
\ee
and, consequently, for the four nonvanishing matrix elements of the potential:
\begin{eqnarray}\nonumber
V_{ik}^{SS}(p,p')&=&\frac{2\pi}{3}\int^1_{-1}F_{ik}(p,p',x)\left(\alpha_{ik} p^2+\alpha'_{ik}p'^2+
(\alpha_{ik} \alpha'_{ik}+1)pp'x\right)dx,\\\nonumber
V_{ik}^{SD}(p,p')&=&-\frac{2\pi\sqrt{2}}{3}\int^1_{-1}F_{ik}(p,p',x)\left
(\alpha'_{ik} p'^2+\alpha_{ik}
p^2\left(\frac32x^2-\frac12\right)+(\alpha_{ik}\alpha'_{ik}+1)pp'x\right)dx,\\\nonumber
V_{ik}^{DS}(p,p')&=&-\frac{2\pi\sqrt{2}}{3}\int^1_{-1}F_{ik}(p,p',x)\left(\alpha_{ik} p^2+\alpha'_{ik}
p'^2\left(\frac32x^2-\frac12\right)+(\alpha_{ik}\alpha'_{ik}+1)pp'x\right)dx,\\\nonumber
V_{ik}^{DD}(p,p')&=&2\pi\int^1_{-1}F_{ik}(p,p',x)\left(\frac23(\alpha_{ik}p^2+
\alpha'_{ik}p'^2)\left(\frac32
x^2-\frac12\right)+\frac{10\alpha_{ik}\alpha'_{ik}+1}{15}pp'x\right.\\
&+&\left.\frac35 pp'\left(\frac52 x^3-\frac32 x\right)\right)dx,
\end{eqnarray}
where $x=\cos\theta$, with $\theta$ being the angle between $\vep$ and $\vep'$. Then, finally, we arrive at the system of four coupled equations:
\begin{eqnarray}
a_{00}^{SS}(p,p',E)&=&V_{00}^{SS}(p,p')\nonumber\\
&-&\int \frac{s^2ds}{\Delta_0(s)}V_{00}^{SS}(p,s)a_{00}^{SS}(s,p',E)
-\int \frac{s^2ds}{\Delta_0(s)}V_{00}^{SD}(p,s)a_{00}^{DS}(s,p',E)\nonumber\\
&-&2\int \frac{s^2ds}{\Delta_c(s)}V_{0c}^{SS}(p,s)a_{c0}^{SS}(s,p',E)
-2\int \frac{s^2ds}{\Delta_c(s)}V_{0c}^{SD}(p,s)a_{c0}^{DS}(s,p',E)\nonumber\\
a_{00}^{DS}(p,p',E)&=&V_{00}^{DS}(p,p')\nonumber\\
&-&\int \frac{s^2ds}{\Delta_0(s)}V_{00}^{DS}(p,s)a_{00}^{SS}(s,p',E)
-\int \frac{s^2ds}{\Delta_0(s)}V_{00}^{DD}(p,s)a_{00}^{DS}(s,p',E)\nonumber\\
&-&2\int \frac{s^2ds}{\Delta_c(s)}V_{0c}^{DS}(p,s)a_{c0}^{SS}(s,p',E)
-2\int \frac{s^2ds}{\Delta_c(s)}V_{0c}^{DD}(p,s)a_{c0}^{DS}(s,p',E)\nonumber\\
a_{c0}^{SS}(p,p',E)&=&2V_{c0}^{SS}(p,p')\label{oureq} \\
&-&2\int \frac{s^2ds}{\Delta_0(s)}V_{c0}^{SS}(p,s)a_{00}^{SS}(s,p',E)
-2\int \frac{s^2ds}{\Delta_0(s)}V_{c0}^{SD}(p,s)a_{00}^{DS}(s,p',E)\nonumber\\
&-&\int \frac{s^2ds}{\Delta_c(s)}V_{cc}^{SS}(p,s)a_{c0}^{SS}(s,p',E)
-\int \frac{s^2ds}{\Delta_c(s)}V_{cc}^{SD}(p,s)a_{c0}^{DS}(s,p',E)\nonumber\\
a_{c0}^{DS}(p,p',E)&=&2V_{c0}^{DS}(p,p') \nonumber\\
&-&2\int \frac{s^2ds}{\Delta_0(s)}V_{c0}^{DS}(p,s)a_{00}^{SS}(s,p',E)
-2\int \frac{s^2ds}{\Delta_0(s)}V_{c0}^{DD}(p,s)a_{00}^{DS}(s,p',E)\nonumber\\
&-&\int \frac{s^2ds}{\Delta_c(s)}V_{cc}^{DS}(p,s)a_{c0}^{SS}(s,p',E)
-\int \frac{s^2ds}{\Delta_c(s)}V_{cc}^{DD}(p,s)a_{c0}^{DS}(s,p',E).\nonumber
\end{eqnarray}

We choose to define the energy $E$ relative to the neutral two-body threshold:
\be
M=m_{0*}+m_0+E.
\ee

The full inverse propagators $\Delta_0$ and $\Delta_c$ entering the system of equations (\ref{oureq}) are obtained as a generalisation of Eq.~(\ref{Deltadef}) through the introduction of the running width $\varGamma(p)$ which incorporates both the effect of the self-energy $\Sigma(p)$ as well as contributions from other $D^*$ decay channels:
\be
\Delta_0(p)=m_{0*}+m_0+\frac{p^2}{2\mu_{0*}}-M-\frac{i}{2}\varGamma_0(p),
\quad
\Delta_c(p)=m_{c*}+m_c+\frac{p^2}{2\mu_{c*}}-M-\frac{i}{2}\varGamma_c(p),
\ee
where the reduced masses are
\be
\mu_{0*}=\frac{m_0m_{0*}}{m_0+m_{0*}},\quad\mu_{c*}=\frac{m_cm_{c*}}{m_c+m_{c*}}.
\label{mus}
\ee

The running widths $\varGamma_0(p)$ and $\varGamma_c(p)$ take the form
\begin{eqnarray}
&&\varGamma_0(p)=\varGamma(D^{*0}\to D^0\gamma)\label{G0}\\
&&+\frac{8\pi^2}{3}g^2\left\{\mu_q(D^0\pi^0)\left[2\mu_q(D^0\pi^0)\left(M-m_0-\frac{p^2}{2\mu_{0*}}-m_0-m_{\pi^0}\right)\right]^{3/2}\right.\nonumber\\
&&\hspace*{9cm}\times\Theta\left(M-m_0-\frac{p^2}{2\mu_{0*}}-m_0-m_{\pi^0}\right)\nonumber\\
&&-i\mu_q(D^0\pi^0)\left[2\mu_q(D^0\pi^0)\left(m_0+\frac{p^2}{2\mu_{0*}}+m_0+m_{\pi^0}-M\right)\right]^{3/2}\nonumber\\
&&\hspace*{9cm}\times
\Theta\left(m_0+\frac{p^2}{2\mu_{0*}}+m_0+m_{\pi^0}-M\right)\nonumber\\
&&+2\mu_q(D^+\pi^-)\left[2\mu_q(D^+\pi^-)\left(M-m_0-\frac{p^2}{2\mu_{0*}}-m_c-m_{\pi^c}\right)\right]^{3/2}\nonumber\\
&&\hspace*{9cm}\times
\Theta\left(M-m_0-\frac{p^2}{2\mu_{0*}}-m_c-m_{\pi^c}\right)\nonumber\\
&&-2i\mu_q(D^+\pi^-)\left[2\mu_q(D^+\pi^-)\left(m_0+\frac{p^2}{2\mu_{0*}}+m_c+m_{\pi^c}-M\right)\right]^{3/2}\nonumber\\
&&\hspace*{9cm}\times
\Theta\left(m_0+\frac{p^2}{2\mu_{0*}}+m_c+m_{\pi^c}-M\right)\nonumber\\
&&\left.+2i\mu_q(D^+\pi^-)\left[2\mu_q(D^+\pi^-)(m_c+m_{\pi^c}-m_{0*})\right]^{3/2}\right\},\nonumber
\end{eqnarray}
and
\begin{eqnarray}
&&\varGamma_c(p)=\frac{8\pi^2}{3}g^2\left\{
2\mu_q(D^0\pi^+)\left[2\mu_q(D^0\pi^+)\left(M-m_c-\frac{p^2}{2\mu_{c*}}-m_0-m_{\pi^c}\right)\right]^{3/2}\right.
\label{Gc}\\
&&\hspace*{9cm}\times
\Theta\left(M-m_c-\frac{p^2}{2\mu_{c*}}-m_0-m_{\pi^c}\right)\nonumber\\
&&-2i\mu_q(D^0\pi^+)\left[2\mu_q(D^0\pi^+)\left(m_c+\frac{p^2}{2\mu_{c*}}+m_0+m_{\pi^c}-M\right)\right]^{3/2}\nonumber\\
&&\hspace*{9cm}\times
\Theta\left(m_c+\frac{p^2}{2\mu_{c*}}+m_0+m_{\pi^c}-M\right)\nonumber\\
&&+\mu_q(D^+\pi^0)\left[2\mu_q(D^+\pi^0)\left(M-m_c-\frac{p^2}{2\mu_{c*}}-m_c-m_{\pi^0}\right)\right]^{3/2}\nonumber\\
&&\hspace*{9cm}\times
\Theta\left(M-m_c-\frac{p^2}{2\mu_{c*}}-m_c-m_{\pi^0}\right)\nonumber\\
&&-i\mu_q(D^+\pi^0)\left[2\mu_q(D^+\pi^0)\left(m_c+\frac{p^2}{2\mu_{c*}}+m_c+m_{\pi^0}-M\right)\right]^{3/2}\nonumber\\
&&\left.\hspace*{9cm}\times
\Theta\left(m_c+\frac{p^2}{2\mu_{c*}}+m_c+m_{\pi^0}-M\right)\right\}.\nonumber
\end{eqnarray}
Note, formally there should be a radiative term also for the decay of the
charged $D^*$, however, in this case it is negligibly small and will therefore
be dropped.

\begin{center}
\begin{figure}[t]
\centerline{\epsfig{file=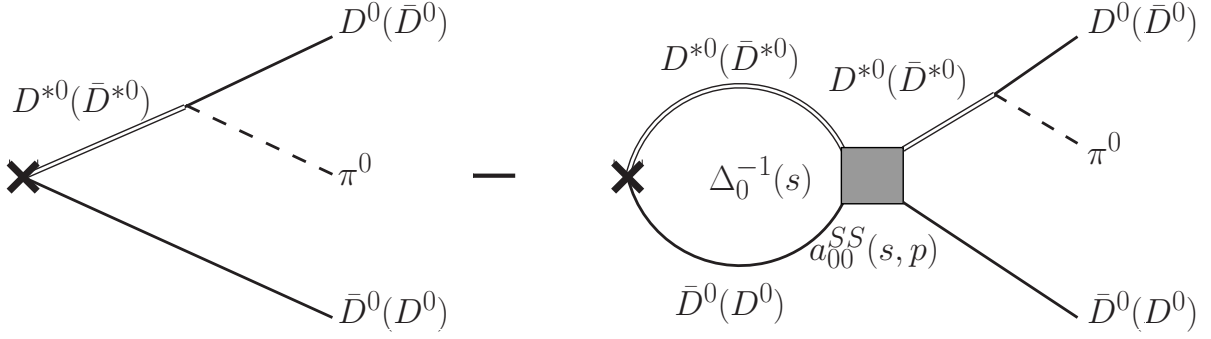, width=16cm}}
\caption{The amplitude for the $D^0 \bar D^0 \pi^0$ $S$-wave production rate. The cross denotes the pointlike source ${\cal F}$.}\label{diagram2}
\end{figure}
\end{center}

The $D^0 \bar D^0 \pi^0$ $S$-wave rate is calculated now as (see Fig.~\ref{diagram2})
\begin{eqnarray}
\label{dbrde}
&&\frac{dBr}{dE}=\frac{\cal B}{2\pi}
\frac{m_{\pi^0}\varGamma_*}{\mu_p(2\mu_q(D^0\pi^0)E_*)^{3/2}}
\int_0^{\sqrt{2\mu_p(E+E_*)}} p dp
\int_{\bar{p}_{\rm min}}^{\bar{p}_{\rm max}} \bar{p} d\bar{p} \nonumber \\
&& \times
\Bigg\{
\left(E+E_*-\frac{p^2}{2\mu_p}\right)
\left| \frac{J(p,E)}{\Delta_0(p)} \right|^2 +
\left(E+E_*-\frac{\bar{p}^2}{2\mu_p}\right)
\left| \frac{J(\bar{p},E)}{\Delta_0(\bar{p})} \right|^2 \\
&& + \frac{1}{\alpha} \left[
\left(\alpha^2+1\right) (E+E_*) -
\frac{p^2+\bar{p}^2}{2\mu_p}
\right]
\mbox{Re} \left[ \frac{J(p,E)}{\Delta_0(p)} \left(
\frac{J(\bar{p},E)}{\Delta_0(\bar{p})}\right)^* \right]
\Bigg\} \nonumber
\end{eqnarray}
where
\beas
&\ds J(p,E)=1-\int_0^{\infty}\frac{s^2ds}{\Delta_0(s)} a_{00}^{SS}(s,p,E),&\\[2mm]
&\ds \bar{p}_{\rm max,min} = \left| \sqrt{2 \mu_q(D^0\pi^0)
\left(E+E_*-\frac{p^2}{2\mu_p}\right) } \pm \alpha p \right|,&\\[2mm]
&\ds E_*=m_{0*}-m_0-m_{\pi^0},\quad
\mu_p=\frac{m_0(m_0+m_{\pi^0})}{2m_0+m_{\pi^0}},\quad
\alpha=\alpha_{00}=\frac{m_0}{m_{\pi^0}+m_0}.&
\eeas
The overall coefficient ${\cal B}=|{\cal F}|^2$ absorbs all the details of the short-ranged dynamics responsible
for the $X$ production.
The last term in the curly brackets in Eq.~(\ref{dbrde}) corresponds to the interference of the production amplitudes
with the pion produced by the $D^{0*}$ and $\bar{D}^{0*}$. In agreement with earlier claims \cite{Vol2004}, for the bound state, interference affects substantially the magnitude of the production rate below the $D^0 \bar D^{0*}$ threshold.

Finally, for numerical calculations, we use the following masses and widths:
\be
\begin{array}{lcl}
m_0=m(D^0)=1864.84~\mbox{MeV},&\quad& m_c=m(D^\pm)=1869.62~\mbox{MeV},\\
m_{0*}=m(D^{*0})=2006.97~\mbox{MeV},&& m_{c*}=m(D^{*\pm})=2010.27~\mbox{MeV},\\
m_{\pi^0}=m(\pi^0)=134.98~\mbox{MeV},&& m_{\pi^c}=m(\pi^\pm)=139.57~\mbox{MeV},
\end{array}
\label{masses}
\ee
\be
\varGamma_*=\varGamma(D^{*0}\to D^0\pi^0)=42~\mbox{keV},\quad \varGamma(D^{*0}\to D^0\gamma)=21~\mbox{keV}. 
\ee

Then, using Eq.~(\ref{D*width}), the coupling constant $g$ can be extracted from the $D^{*0}\to D^0\pi^0$ width to be:
\be
g= 1.29\cdot 10^{-5}~\mbox{MeV}^{-3/2}.
\ee

\subsection{Regularisation and renormalisation of the three-body equation}

The system (\ref{oureq}) is to be solved numerically.
Because of the $P$-wave $D^*D\pi$ vertex [see
Eq.~(\ref{g1})], the integrals on the right-hand side diverge linearly.
We separate the short- and long-range
dynamics of the system, where the long-range interaction is due to the
OPE while the short-range one, in addition to the short-range
part of the OPE, may contain other contributions, for example, due to a strong
coupling of the $D \bar D^*$ to the quark--antiquark charmonium. This
short-range dynamics is parametrised by a
constant (in momentum space) $C_0(\Lambda)$ (below referred to as contact
interaction or as counter term) which appears just as an extra term in the
potential $V_{ik}^{SS}$, that is now \be
V_{ik}^{SS}(p,p')=C_0(\Lambda)+\frac{2\pi}{3}\int^1_{-1}F_{ik}(p,p',x)\left(\alpha_{ik}
p^2+\alpha'_{ik}p'^2+ (\alpha_{ik} \alpha'_{ik}+1)pp'x\right)dx,
\label{Vs}
\ee while all other components of the potential are left intact. Here
$\Lambda$ is the cut off parameter which regularises the integrals in
Eq.~(\ref{oureq}) (we use the simplest sharp cut off prescription, that is we
substitute $\int_0^\infty ds\to\int_0^\Lambda ds$). Ideally one chooses for
every $\Lambda$ the value of the counter term in such a way that it
neutralises the dependence of the physical observables near threshold on the cut off
$\Lambda$. As we will discuss below, this does not work for some values of
$\Lambda$ at least for the particular regularisation used to solve the system (\ref{oureq}).
This may indicate the need of additional counter terms.
However, since the residual $\Lambda$ dependence is very mild
in a large range of cut offs ($300~\mbox{MeV}\lesssim\Lambda\lesssim 1700$~MeV and
$2500~\mbox{MeV}\lesssim\Lambda\lesssim 3800$~MeV), the renormalisation with a
single counter term appears to be appropriate to investigate the issues at
hand---the complete discussion of the properties of the system (\ref{oureq}) will
be postponed to a subsequent publication.

\section{The effect of dynamical pions on a near-threshold resonance}

We will now compare the solution
of the full problem (\ref{oureq}) (including the counter term discussed
in the previous chapter), with the solution of the same problem in the static
approximation. This implies using the static OPE potential
and omitting the momentum dependence of the ${D^*}^0$ width, cf.~Sec.~\ref{Rate}
and Eq.~(\ref{tildeD3staticdef}) for more details. In both
calculations we require the system to possess a resonance at $E=-E_B$ with
$E_B=0.5$~MeV. Note that in the presence of pions there is no bound state, so we define 
the position of the resonance as the position of a peak in the $D^0{\bar D}^0 \pi^0$ production
rate, given by Eq.~(\ref{dbrde}). 
Therefore, comparing the properties of the results of the
two calculations we investigate the role played by dynamical pions. In
addition we study the role played by the charged channel in the formation of
the $X(3872)$ state.

\subsection{Investigation of the $D \bar D^*$ contact interaction}\label{ci}

Before we proceed, we briefly discuss the calculations for
a model with the contact $D\bar{D}^*$ interaction only [just retaining the $C_0$ term in Eq.~(\ref{Vs})].
The advantage of this model is that it is solvable analytically and thus
a comparison of this simplified model with the results of the full calculation
allows one to understand better the findings.
In the next subsections the corresponding equations for the single-channel as well as for
the coupled-channels problem are derived, for the explicit solution will allow us
to better understand some  properties of the full equations.

\subsubsection{Single-channel case with a contact $D \bar D^*$ interaction}\label{Sci}

We start from the simplest, single-channel, case of Eq.~(\ref{oureq}) with the
only nonvanishing potential being $V_{00}^{SS}=C_0$. Then
Eq.~(\ref{oureq}) reduces to a simple algebraic equation:
\be 
a_{00}=C_0-C_0a_{00} I_0,
\label{LSE1}
\ee
where, in order to simplify notations, we set in this chapter $a_{00}\equiv a_{00}^{SS}$. Here
\be
I_0(E)=\int_0^{\Lambda} ds \frac{s^2}{s^2/(2\mu_{0*})-E-i0}=2\mu_{0*} \left(\Lambda + \frac{k}{2}\left(i\pi+
\log\frac{\Lambda-k}{\Lambda+k}\right)\right),
\ee
with $k^2=2\mu_{0*} E$. 

The strength of the contact interaction $C_0$ is fixed by the requirement that the system possesses a bound state at $E=-E_B$ ($E_B>0$):
\be
C_0^{-1}=-I_0(-E_B),\quad a_{00}^{-1}=I_0(E)-I_0(-E_B).
\ee
To proceed we remind the reader that Eq.~(\ref{LSE1}) is nothing but a toy model for the full system (\ref{oureq}) 
where the cut off $\Lambda$ is to be chosen large enough, that is larger than the natural scales in the full dynamical problem. 
This implies that $\Lambda\gg\sqrt{\mu_{0*}E_B}$ at least, so that the integral $I_0$ can be expanded in powers of the ratio $k/\Lambda$, 
\be
I_0\approx 2\mu_{0*} \left(\Lambda +\frac{i}{2}\pi k-\frac{k^2}{\Lambda}\right) +O\left(\frac{k^4}{\Lambda^3}\right),
\label{I0}
\ee
which corresponds to the effective-range expansion of the scattering amplitude:
\be
a_{00}^{-1}=-\mu_{0*}\pi\left(-a^{-1}-ik+\frac12 r_{\rm eff}k^2\right),
\label{a002}
\ee
where the scattering length and the effective range take the form:
\be
a=\frac{1}{\sqrt{2\mu_{0*} E_B}}\frac{1}{1-\sqrt{2 \mu_{0*} E_B}r_{\rm eff}/2},\quad r_{\rm eff}=\frac{4}{\pi \Lambda}.
\ee

The amplitude $a_{00}$ possesses two poles in the complex $k$ plane:
\be
k_1=i\sqrt{2 \mu_{0*} E_B},\quad k_2=i\left(\frac{2}{r_{\rm eff}}-\sqrt{2 \mu_{0*} E_B}\right).
\ee

In particular, for $E_B=0.5$~MeV and $\Lambda=500$~MeV, one finds
\be
a=6.6~\mbox{fm},\quad r_{\rm eff}=0.5~\mbox{fm},\quad k_1=i31.1~\mbox{MeV},\quad k_2=i754.3~\mbox{MeV}.
\label{nums1}
\ee

\begin{figure}
\centering
\includegraphics[width=10cm]{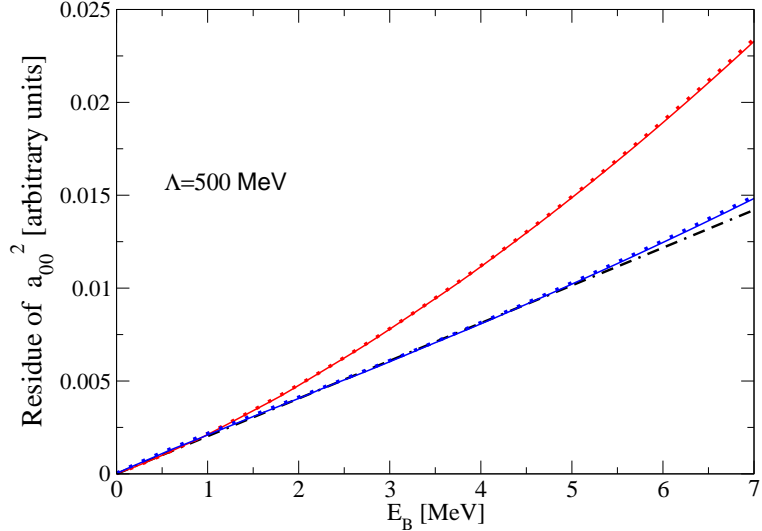}
\caption{Residue of the $D^0\bar{D}^{*0}$ scattering amplitude squared
versus the binding energy in the $D^0\bar{D}^{*0}$ system. The upper, red
(lower, blue) dotted curve corresponds to the solution of the single(two)-channel
$D^0\bar{D}^{*0}$ problem with the contact $D \bar D^*$ interaction. Solutions of the
full three-body equation with dynamical pions are given by the solid
lines: upper, red line---for the single-channel case and lower, blue line---for the
two-channel case. The straight dot-dashed line (black) is shown to guide the
eye.}
\label{res}
\end{figure}

Finally, expanding $k$ at the bound-state energy for $|E+E_B|\ll E_B$,
\be
k=\sqrt{2\mu_{0*} E}\approx i\sqrt{2\mu_{0*} E_B}\left(1-\frac{E+E_B}{2E_B}\right),
\ee
we arrive at the expansion of the amplitude $a_{00}$ at the pole:
\be
a_{00}=\frac{\mbox{Res}~ a_{00}}{E+E_B}\equiv\frac{g_{\rm eff}^2}{E+E_B},
\ee
where the residue reads
\be
\mbox{Res }a_{00}=
\frac{\kappa_0(E_B)}{\mu_{0*}^2\pi}\left(1-\frac{4}{\pi\Lambda}\kappa_0(E_B)\right)^{-1}=
\frac{\sqrt{2\mu_{0*} E_B}}{\mu_{0*}^{2}\pi(1-r_{\rm eff}\sqrt{2\mu_{0*} E_B})},
\label{resa002}
\ee
with $\kappa_0(E_B)=\sqrt{2\mu_{0*} E_B}$.
This residue squared is shown in Fig.~\ref{res} for $\Lambda=500$~MeV as the red (upper) dotted line.
Its mild deviation from the straight line, at least for $E_B<3$ MeV, 
is caused by a nonzero but small value of the effective range, so that we have, approximately,
\be
g_{\rm eff}^2\propto \sqrt{E_B}.
\label{geff}
\ee

Summarising, one can see that we have a large scattering length and one
near-threshold pole in the $k$ plane. The residue squared (that is $g_{\rm eff}^4$) of the amplitude at
the bound-state pole scales almost linearly with the bound-state energy.
These results allow for a nice physical interpretation using the method
suggested by S.~Weinberg in the mid 60's \cite{Weinberg}. He showed at the
example of the deuteron, that an analysis of low-energy observables allowed one
to quantify, in terms of the probability factor $Z$ ($0\le Z\le 1$), the
admixture of the bare (``elementary'') state in the wave function of a
near-threshold bound state. This approach was generalised in Ref.~\cite{evi} to the
case when inelastic channels are present, as well as to the case of an
above-threshold resonance. In particular, in Ref.~\cite{evi}, the connection is
established between the admixture of a bare state and the structure of
near-threshold singularities of the scattering amplitude. The latter was
considered in Ref.~\cite{Morgan} as a tool for resonance classification. The
results of Refs.~\cite{evi,Morgan} can be briefly summarised as follows. A
state is mostly elementary, if there are two nearby poles in the scattering
amplitude, which corresponds to a small scattering length, and large and
negative effective range. In this case the coupling of the state to the
hadronic channel is small, and the resonance line shape takes a Breit--Wigner
form. The state is mostly composite, if the scattering length is large, and
there is only one near-threshold pole. The effective range is of natural size
(of the order of range of forces) and plays a role of a correction. This
scenario requires a large coupling of the state to the hadronic channel.
In this case the wave
function is dominated by its molecular component. The residue of the relevant
pole is determined by the binding energy of a composite particle and thus it
is model-independent.
For the parameters used as an illustrative example above [see Eq.~(\ref{nums1})]
we clearly have a predominantly molecular state.

\subsubsection{Two-channel case with contact $D \bar D^*$ interactions}\label{Tci}

In the case of two coupled channels, the scattering equation for the two
components of the amplitude, for the sake of simplicity denoted in this chapter as $a_{00}\equiv a_{00}^{SS}$ and
$a_{c0}\equiv a_{c0}^{SS}$, takes the form: 
\be 
\left\{
\begin{array}{rcl}
a_{00}&=&C_0-C_0 a_{00}I_0-2C_0 a_{c0} I_c\\
a_{c0}&=&2C_0-2C_0 a_{00} I_0 - C_0 a_{c0} I_c,
\end{array}
\right.
\ee
with the solution for $a_{00}$ given by
\be
a_{00}=\frac{C_0(1-3C_0I_c)}{(1+C_0 I_0)(1+C_0 I_c)-4 C_0^2 I_0I_c}.
\label{a00couple}
\ee
Similarly to the single-channel case, the integrals are
\bea
I_0&=&\int_0^{\Lambda} ds \frac{s^2}{s^2/(2\mu_{0*})-E-i0}\approx 2\mu_{0*} \left(\Lambda +\frac{i}{2}\pi k_0-\frac{k_0^2}{\Lambda}\right) +O\left(\frac{k_0^4}{\Lambda^3}\right),\\
I_c&=&\int_0^{\Lambda} ds \frac{s^2}{s^2/(2\mu_{c*})+\Delta M-E-i0}\approx 2\mu_{c*} \left(\Lambda +\frac{i}{2}\pi k_c-\frac{k_c^2}{\Lambda}\right) +O\left(\frac{k_c^4}{\Lambda^3}\right), \label{Ic}
\eea
where $k_0^2=2\mu_{0*} E$, $k_c^2=2\mu_{c*} (E-\Delta M)$, and $\Delta M =m_c^*+ m_c -m_0^* -m_0= 8.08$~MeV.
As before, we require that there is a bound state at $E=-E_B$, that is the amplitude (\ref{a00couple}) has a pole at this point. This leads to the equation to determine the strength of the contact term $C_0$. Notice that now, in the two-channel case, this equation is quadratic, as opposed to the linear equation in the single-channel case:
\be
\Bigl(1+C_0I_0(-E_B)\Bigr)\Bigl(1+C_0I_c(-E_B)\Bigr)-4 C_0^2I_0(-E_B)I_c(-E_B)=0.
\label{pole}
\ee
It is easy to verify that the latter equation always possesses two opposite-sign solutions for the $C_0$. In particular, for $\Lambda=500$~MeV and $E_B=0.5$~MeV these solutions are $C_0=13.442\times 10^{-7}$~MeV$^{-2}$ and
$C_0=-4.425\times 10^{-7}$~MeV$^{-2}$.

Notice that the coupled-channel problem is still dominated by one relevant
model-independent near-threshold pole which, for $E_B=0.5$~MeV and
$\Lambda=500$~MeV, is still located at $k_1=i31.1$~MeV equal to the value of $k_1$
for the single-channel case given in Eq.~(\ref{nums1}). The position of other
poles is very model-dependent, however they all are in general located quite far away
from the threshold.

Following the same lines as in the single-channel case above, one can find the
residue of the amplitude $a_{00}$ at $E=-E_B$. The dependence of the residue
squared on the binding energy for $\Lambda=500$~MeV is illustrated in
Fig.~\ref{res} by the lower (blue) dotted line. To a good approximation, the
dependence of the residue on the binding energy can be written as
\be {\rm
Res}~a_{00}\approx\frac{\mbox{const($\Lambda$)}}{\pi\mu_{0*}^2}{\frac{\kappa_c(E_B)\kappa_0(E_B)}{\kappa_c(E_B)+\kappa_0(E_B)}}
\left(1-\frac{8}{\pi\Lambda}\frac{\kappa_c(E_B)\kappa_0(E_B)}{\kappa_c(E_B)+\kappa_0(E_B)}\right)^{-1},
\label{resa003}
\ee
where $\kappa_c(E_B)=\sqrt{2\mu_{c*} (E_B+\Delta M)}$, $\kappa_0(E_B)=\sqrt{2\mu_{0*} E_B}$,
and const($\Lambda$) is a constant of order of unity.
In the limit of small binding energies ($E_B\ll \Delta M$), one has $\kappa_0(E_B)\ll \kappa_c(E_B)$, and the
formula for the residue (\ref{resa003}) can be simplified as
\begin{eqnarray}
{\rm Res }~a_{00}\approx\mbox{const}\cdot \frac{\kappa_0(E_B)}{\pi\mu_{0*}^2}\left(1-\frac{8}{\pi\Lambda}\kappa_0(E_B)\right)^{-1},
\end{eqnarray}
so that, similarly to the single-channel case [compare with
Eq.~(\ref{resa002})], the residue squared remains approximately linear with
$E_B$, up to small finite-range corrections.
In principle, the range of validity of this expression is rather limited since
the correction $\sim \sqrt{E_B}$ from $\kappa_c(E_B)$ starts to play a role quite rapidly. However,
an analogous correction but with an opposite sign appears also from the finite-range term
($\sim 1/\Lambda$) in the denominator. This leads to a partial cancellation of these terms.

Interestingly, for $E_B\sim \Delta M$, the linear behaviour of the residue
squared as a function of $E_B$ is still approximately preserved since again the
finite-range corrections are cancelled to a large extent by the corrections
due to $\kappa_c(E_B)$. As a result, for $\Lambda=$500~MeV, the behaviour of
the residue squared is almost exactly linear in the whole range of $E_B$. This
linearity is, of course, accidental---for larger cut offs the deviation from
the straight line is more pronounced; however, it remains a correction to the
leading linear behaviour.

Deviations from the Weinberg prediction are expected to be of order
$r{\sqrt{2\mu_{0*}E_B}}$, with $r$ being the range of forces. If we estimate the latter by $r_{\rm eff}$
from above, $r_{\rm eff}{\sqrt{2\mu_{0*}E_B}}\sim 0.1\sqrt{E_B/{\rm MeV}} $,
we expect a deviation of the residue from the Weinberg prediction of the
order of 30\% at $E_B=7$~MeV, in line with what can be read off from Fig.~\ref{res}
for the single-channel curve. As explained above, accidental cancellations
make the result for the coupled-channel calculation more consistent
with a linear behaviour than expected naively.

The influence of the coupled-channel dynamics on the residue can be better understood if
one considers the limit
$\Lambda\to \infty$. In this case, as follows from Eqs.~(\ref{resa002}) and (\ref{resa003}),
the ratio of the residue for the coupled-channel problem to that for the single-channel one reads
\be
R=\frac{{\rm Res}~a_{00}(\mbox{coupled-channel})}{{\rm Res}~a_{00}(\mbox{single-channel})}
\mathop{=}\limits_{\Lambda\to\infty}\left(1+\sqrt{\frac{E_B}{E_B+\Delta M}}\right)^{-1},
\label{R}
\ee
where we neglected the tiny difference between $\mu_{0*}$ and $\mu_{c*}$ and
used the fact that const($\Lambda)\to 1$ when $\Lambda\to \infty$.
It is clear from the ratio (\ref{R}) that the coupled-channel effect is negligible at small binding energies, $E_B\ll\Delta M$, that is, once we stay close to one threshold, its effect dominates over the effect of the other, remote, threshold. In the opposite limit, for $E_B\gg\Delta M$, when the splitting between thresholds can be neglected, this ratio tends to one half, that is, in agreement with natural expectations, asymptotically both thresholds contribute equally to the residue.

\subsubsection{Counter-term in presence of dynamical pions}

\begin{center}
\begin{figure}[t]
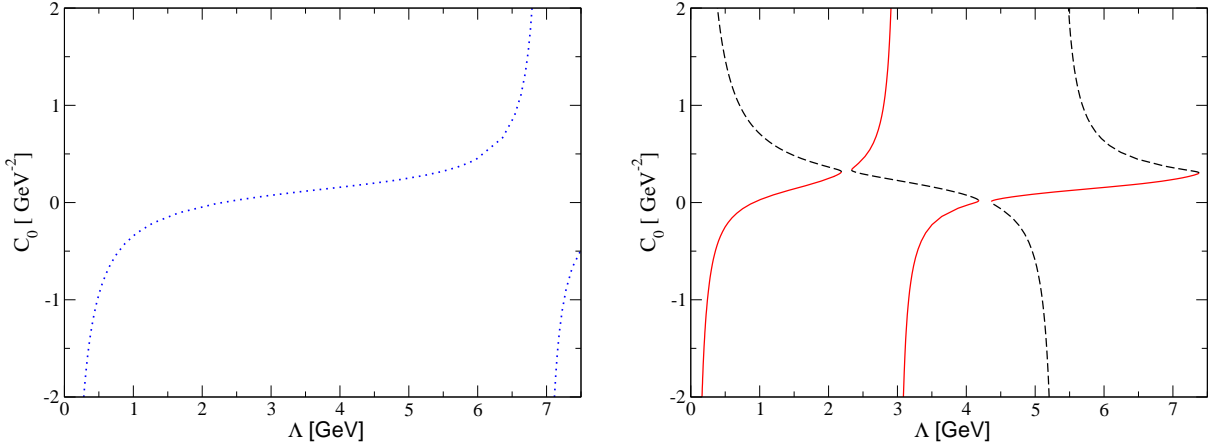

\epsfig{file=limit_cycle_1chan.eps, width=7.7cm,clip=}\hspace*{0.5cm}
\epsfig{file=limit_cycle_2chan.eps, width=7.7cm,clip=}
\caption{Left panel: Behaviour of the counter term $C_0$ versus $\Lambda$ in the full single-channel problem ($D^0{\bar D}^{0*}$) with dynamical pions. Right panel: Behaviour of  $C_0$ as a function of the cut off $\Lambda$ in the full two-channel problem with dynamical pions. }\label{c0fig}
\end{figure}
\end{center}

The behaviour of the counter term $C_0$ versus $\Lambda$ in the full single-channel problem ($D^0{\bar D}^{0*}$)
with dynamical pions is shown  in the left panel of Fig.~\ref{c0fig} by the blue dotted curve.
The figure demonstrates a clear limit-cycle behaviour
of the contact term with the increase of $\Lambda$  in full analogy with the NN \cite{Nogga}
and 3N \cite{Bedaque} problems, see also Refs.~\cite{Kud,Beane,Braaten} for related works.
Meanwhile, a plateau between the first negative and positive  infinite solutions  is about 7 GeV  
for the $D{\bar D}^*$ problem which is much larger than that for NN~\cite{Nogga}. 

The behaviour of the counter term $C_0$ on the cut off $\Lambda$ in the two-channel
case, when dynamical pions are present in the problem, is illustrated in the right panel of Fig.~\ref{c0fig}.
We therefore solve the full problem---see Sec.\ref{fullsec}.  
In contradistinction to the one-channel case 
(and to the NN case), in a  coupled-channel problem there are two solutions for $C_0(\Lambda)$ 
that correspond  to a resonance with the peak at $E_B=0.5$ MeV [see the quadratic equation (\ref{pole}) 
and the discussion in the previous section]. These solutions, shown as solid red and
dashed black curves in the right panel of Fig.~\ref{c0fig}, again exhibit a sort of limit-cycle behaviour individually. 
However, it is interesting to observe that there are regions of cut offs where no solution
for $C_0(\Lambda)$ exists. This can be understood as follows: for values of $\Lambda$,
where the two solutions for $C_0(\Lambda)$
approach each other (see, for example, the region of $\Lambda$ just above 2~GeV),
a second singularity approaches the $D\bar{D}^*$ threshold. As a consequence, the two approaching states
start to repel each other to avoid a crossing of the levels. Therefore there is a region of  $E_B$ where the peak 
of the resonance  cannot be reached
by any variation of the strength of the potential $C_0(\Lambda)$. The regions of $\Lambda$ corresponding to the
absence of the resonance peak at $E_B=0.5$ MeV for any values of $C_0(\Lambda)$ are illustrated in the right 
panel of Fig.~\ref{c0fig}.
Further details on the limit cycle behaviour in many-channel problems will be provided in a subsequent publication.

What should be stressed here is that as long as we choose the cut off to be
relatively far away from the problematic region one of the states is always located far away from the threshold
and therefore should not affect observables. As a consequence,
observables are basically independent of whether one or the other branch of $C_0$ is chosen---see Sec.\ref{Rate}.

A comment is in order here concerning the role played by the OPE for the binding of the $X$ meson.
From Fig.~\ref{c0fig} one can see that for particular values of the cut off 
(for example, for $\Lambda\approx 900$ MeV in the coupled-channel problem),  
there is a solution with $C_0=0$ which could be naively taken for the bound 
state in the pure OPE potential with dynamical pions. This is not the case 
however since the details of the short-ranged dynamics are simply hidden in 
the particular cut off pattern for the divergent integrals taking place for 
such $\Lambda$'s. Furthermore, in previous chapters, we showed how to arrive 
at the bound state with a purely contact potential. Finally, the bound state 
can be found in a combined short-range+OPE potential with $C_0$ varying from 
$-\infty$ to $\infty$, as shown in Fig.~\ref{c0fig}. Since physical observables, 
for example, the width of the $X$-meson (see below) are almost $\Lambda$-independent 
in a large interval of $\Lambda$'s, then no model-independent statement is possible 
concerning the importance of the one-pion exchange for the
formation of the $X(3872)$, contrary to earlier claims~\cite{tornqvist2,ThCl}.

\subsection{The effect of dynamical pions on a near-threshold resonance}
\label{Rate}

With the experience gained in the previous chapter, we are in a position to
perform a detailed numerical analysis of the system (\ref{oureq}). We compare
the following three situations:
\begin{enumerate}
\item The single-channel problem in the static approximation, that is we solve the system (\ref{oureq}) with
$V_{c0}^{mn}(\vep,\vep')=V_{0c}^{mn}(\vep,\vep')=V_{cc}^{mn}(\vep,\vep')=0$
and with the static OPE potential which,
as before, is given by Eq.~(\ref{Vmn2}), but with $\alpha_{00}=\alpha'_{00}=1$ and with
\be
D^{\rm static}_{300}(\vep,\vep')=2m_0+m_{\pi^0}+\frac{(\vep+\vep')^2}{2m_{\pi^0}}-(m_{0*}+m_0)-i0
\label{tildeD3staticdef}
\ee
for the inverse three-body propagator. In addition, instead of the running width (\ref{G0}), we use a constant width
\be
\varGamma_0=63~\mbox{keV}.
\ee
Note that in the absence of the three-body dynamics both these effects, the constant width and the imaginary part of the potential,
represent the same effect of the decay $D^* \to D\pi$ related to two-body unitarity. Thus, keeping only one of these
effects would lead to inconsistent treatment.

\item The full dynamical calculation for the single-channel problem,
including the three-body $\pi D\bar D$ intermediate states as well as the dynamical width of the $D^*$.
For the $D^{*0}$ width we use formula (\ref{G0}) without the contribution of
the charged channels.
\item The full dynamical calculation of the two-channel problem,
including three-body $D\bar D\pi$ intermediate states as well as the dynamical width of the $D^*$, as was explained in Sec.\ref{fullsec}.
\end{enumerate}

\begin{center}
\begin{figure}
\centerline{\epsfig{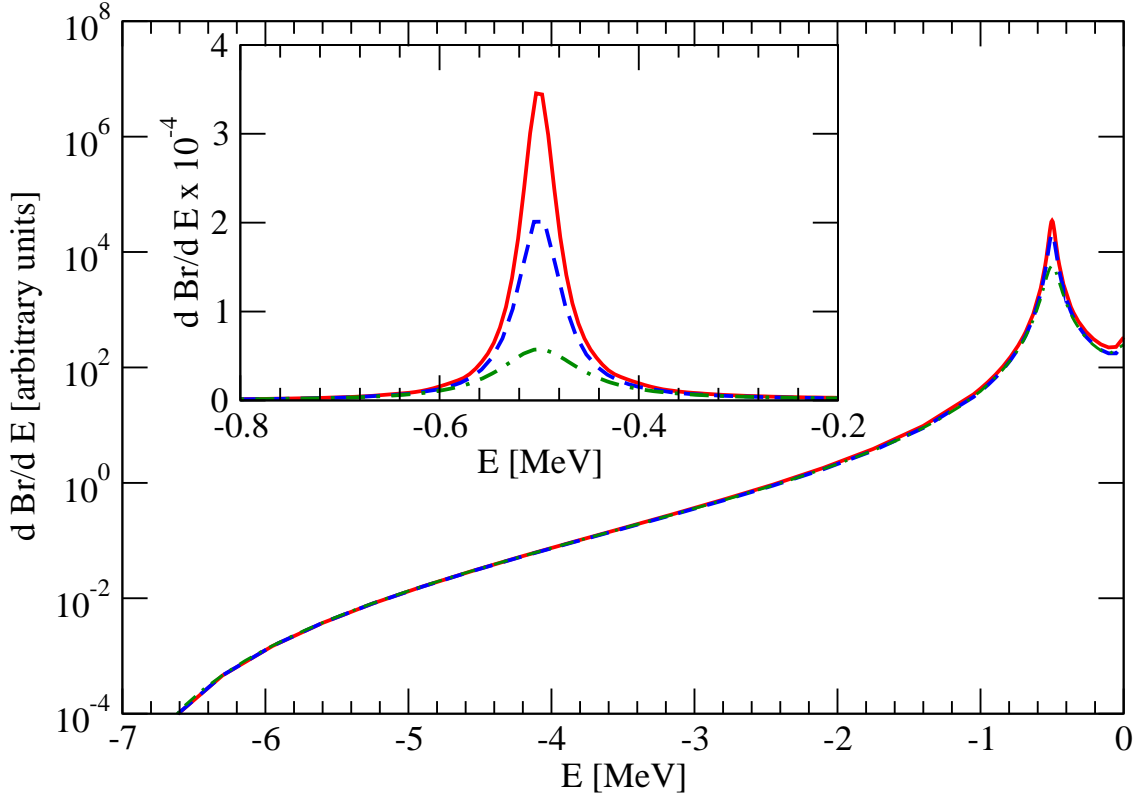}}
\caption{Production rate (in logarithmic scale) for the three calculations as
described in the text: (i) solution of the single-channel problem in the
static limit---(green) dot-dashed line; (ii) solution of the single-channel
dynamical calculation---(blue) dashed line; (iii) solution of the full
two-channel dynamical problem---(red) solid line. All curves are normalised
near the $D^0\bar D^0\pi^0 $ threshold, located at $E=-7$ MeV. The
inlay shows a zoom into the peak region in linear scale. }
\label{rate}
\end{figure}
\end{center}

First, we investigate the effect of dynamical pions on the residue of the scattering amplitude $a_{00}$
squared---see Fig.~\ref{res} and the discussion in Sec.\ref{ci}.
The upper (red) and lower (blue) solid curves correspond to case 2 and case 3 above, respectively. As was discussed before,
the solid curves lie on top of the dotted curves indicating that the binding energy dependence of the residue
is basically independent of the dynamics. The closeness of the results to the straight line 
(black dot-dashed curve in Fig.~\ref{res})
indicates the dominance of the single-pole scenario and thus the molecular nature of the $X(3872)$ meson.

In Fig.~\ref{rate} we show the results for the production rate $d Br/d E$
corresponding to all three different cases described above: (i) solution of
the single-channel problem in the static limit---(green) dot--dashed line; (ii)
solution of the single-channel dynamical calculation---(blue) dashed line; (iii)
solution of the full two-channel dynamical problem---(red) solid line. In all
cases the value of the cut off $\Lambda$ was fixed to 500 MeV and
the strength of the contact operator $C_0$ was adjusted such 
that the $D^0\bar{D}^{*0}$ scattering amplitude has a resonance state peaked at
$E=-E_B=-0.5$~MeV. All curves are normalised to obey the same energy behaviour near
the three-body $D^0\bar D^0\pi^0 $ threshold, located at $E=-7.15$ MeV. The difference between them is
visible only in the region of the peak  and is pronounced in the different
strength of the peaks---see Fig.~\ref{rate}. 
If one normalises all three curves at the peak, the difference in the resonance region shows up only in the
different widths of the resonance. To quantify the effect we notice that, in the energy region around the resonance peak,
the $D\bar D^*$ scattering amplitude $a_{00}$ acquires basically a separable form:
\be
a_{00}^{SS}(\vep,\vep',E)=-\frac{1}{\mu_{0*}\pi}f(E) \phi(\vep) \phi(\vep'),
\ee
with $\phi$ being a formfactor,
so that the energy-dependent amplitude $f(E)$ is factored out from 
the integration over the phase space in Eq.~(\ref{dbrde}). Parametrising 
$f(E)$ by the Breit--Wigner (BW) shape, one finds for the rate:
\be
\left(\frac{dBr}{dE}\right)_{\rm BW}=\frac{\mbox{const}}{(E+E_B)^2+\varGamma_X^2/4} {~\hat{k}_{\rm eff}(E)},
\label{BW}
\ee
where $\hat{k}_{\rm eff}(E)$ corresponds to the integral over the phase space in Eq.~(\ref{dbrde}) with $J(p,E)=1$.
Except for the region very close to the threshold, $\hat{k}_{\rm eff}(E)$
is a very smooth function of energy around the peak as compared to the Breit--Wigner shape. Therefore,
expression (\ref{BW}) can be used to extract the width $\varGamma_X$.  
The results for $\varGamma_X$ extracted for all three cases above are shown in Table~\ref{fit}. 
Examining the transition from case 2 to case 3, one can see  that  including the charge
channel changes the shape of the spectrum only marginally. The nearest threshold which  affects the
observables due to the coupled-channel effect is $D^+\bar D^0\pi^- $ (and $D^- D^0\pi^+$)
which is about 2.5 MeV above the $D^0\bar{D}^{*0}$ threshold. Also the $D^+ D^-\pi^0 $ threshold is just 3~MeV above
the $D^0\bar{D}^{*0}$ threshold.
Had the mass of the charged $D$-meson been smaller by $2.5-3$ MeV and more, the new cuts would have occurred
in the $D\bar{D}^{*}$ potential increasing the $X(3872)$ width and thus emphasising the role
of the coupled-channel dynamics. In the meantime,
the transition from case 1 to case 2 elucidates the role of the three-body
dynamics: switching on three-body effects makes the resonance narrower by a factor
of 2. 

In order to quantify the role of the pion dynamics in the vicinity of the threshold we use the scattering length approximation and study the impact of the three-body dynamics on
the $D^0\bar{D}^{*0}$ scattering length. The molecule scenario for the $X(3872)$ implies that, near threshold, the effective-range corrections are small 
(which is also in agreement with the nearly linear behaviour of the residue as a function of the binding
energy in Fig.~\ref{res}), 
so we resort to the scattering length (SL) approximation for the production rate:
\be
\left(\frac{dBr}{dE}\right)_{\rm SL}=\mbox{const}\times|f(E)|^2 {~\hat{k}_{\rm eff}(E)},
\quad f(E)=\frac1{-\gamma_1-i\gamma_2-i\sqrt{2\mu_{0*}(E+i0)}}.
\label{scatt}
\ee

A similar approach to the analysis of the $DD\pi$ line shape was used in Ref.~\cite{stapleton}, where the width of the $X$ originated from a constant $D^{0*}$ width in the absence of inelastic channels. In Eq.~(\ref{scatt}) we also neglect inelastic effects from remote thresholds and
introduce $\gamma_2$ to account for the change of the $X$ width 
due to the $D\bar{D}\pi$ intermediate states in the energy region near the $D\bar D^*$ threshold. 
Stated differently, $\gamma_2$ accounts for the shift 
of the pole position in the complex plane due to the three-body dynamics. 
Parameters $\gamma_1$ and $\gamma_2$ ($\gamma_2 > 0$ from unitarity) corresponding to
the full dynamical treatment can be well determined from a fit to the resonance structure
in the line shape---see Fig.~\ref{rate}. 

Near the peak, that is for $|E+E_B|\ll E_B$, the scattering length
formula (\ref{scatt}) is identical to the Breit--Wigner one (\ref{BW}) with the following identification of the parameters:
\be
E_B=\frac{\gamma_1^2-\gamma_2^2}{2\mu_{0*}}, \quad \varGamma_X=\frac{2\gamma_1\gamma_2}{\mu_{0*}}.
\ee
Furthermore, for a narrow resonance ($\varGamma_X\ll E_B$), these relations can be inverted to give:
\be
\gamma_1\approx \sqrt{2\mu_{0*} E_B},\quad \gamma_2\approx \frac{\mu_{0*} \varGamma_X }{ \sqrt{2\mu_{0*} E_B}}.
\label{gamma2}
\ee
Clearly, a reduction of the $X$ width by a factor of 2 would necessarily suppress $\gamma_2 $ by the same amount.
The results for $\gamma_1$ and $\gamma_2$ for three different scenarios from above are quoted in Table~\ref{fit}.
It should be noted that the expression (\ref{scatt})
provides a very good approximation for the exact line shape in the near-threshold region even relatively 
far away from the pole, for $|E+E_B|\sim E_B$. 
\begin{table}
\begin{ruledtabular}
\begin{tabular}{lcccc}
Case & $E_B$, MeV& $\varGamma_X$, keV& $\gamma_1$, MeV & $\gamma_2$, MeV\\
\hline
Case 1 &        0.5  & 102 & 31.1 & 1.58\\ 
Case 2 &        0.5  & 53 & 31.1 & 0.82\\ 
Case 3 &        0.5  & 44 & 31.1 & 0.68\\ 
\end{tabular}
\end{ruledtabular}
\caption{Parameters of the distributions (\ref{BW}) and (\ref{scatt}) extracted from the fit to the peak for the line shapes depicted in Fig.~\ref{rate}.}\label{fit} 
\end{table}

\begin{center}
\begin{figure}
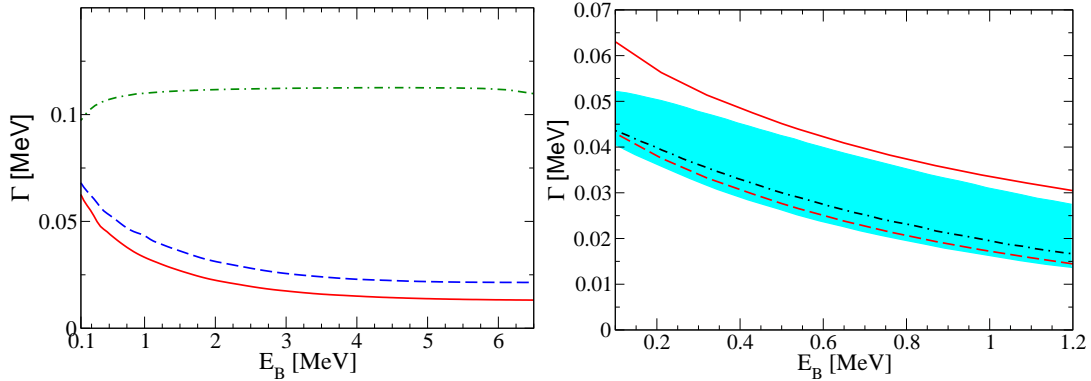

{\epsfig{file=width_Ebpaper.eps, width=7cm}
\epsfig{file=width_comparison.eps, width=7.2cm}}
\caption{The $X$ width as a function of the binding energy $E_B$.  Left panel:
The results of three calculations as explained in the text; the notation of
curves is as in Fig.~\ref{rate}. Right panel: Comparison of our results [red
solid (dashed) curve including (omitting) the  $\bar D^* D\gamma$ width] with those of Refs.~\cite{Vol2004,pions} (dot--dashed line for the LO calculation of Ref.~\cite{Vol2004} and blue band for the NLO
calculation of Ref.~\cite{pions}, respectively). }
\label{width}
\end{figure}
\end{center}

We would like to stress also that the effect of pion dynamics depends strongly
on the position of the resonance state relative to the relevant two- and
three-body thresholds. Indeed, once it approaches the three-body $D^0\bar
D^0\pi^0 $ threshold, the imaginary parts originating from the three-body
effects vanish, and the $X$ width tends to a constant due to the $D^{0*}\to
D^0\gamma$ transition, while the $X$ width in the static approximation stays
nearly constant---see left panel of Fig.~\ref{width}. On the other hand,
the dynamical width of the $X(3872)$ grows as the resonance peak approaches
the two-body $D^0\bar D^{*0}$ threshold. The dependence of the effective $X$
width $\varGamma_X$ [as well as that of the $\gamma_2$ by virtue of
Eq.~(\ref{gamma2})] on the binding energy can be parametrised in the interval
from 0.1 to 2~MeV in an analytic form: \be
\varGamma_X(E_B)=\frac{\varGamma_1\beta_1^2}{E_B^2+\beta_1^2}+\frac{\varGamma_2\beta_2^2}{E_B^2+\beta_2^2},
\ee with $\varGamma_1=35.5$~keV, $\beta_1=2.369$~MeV, $\varGamma_2=28.7$~keV,
and $\beta_2=0.364$~MeV. In this form the effect of dynamical pions can be
conveniently incorporated in the analysis of experimental data.

In the right panel of Fig.~\ref{width} the results of our full dynamical
calculation are confronted with those of the $X$-EFT \cite{Vol2004,pions}.  In
leading order (LO) of $X$-EFT \cite{Vol2004} pion degrees of freedom are
integrated out, and the wave function of the $X$, as a bound state, is taken
in a universal form provided by the contact $D\bar D^*$ interaction. Then the
$X$ width is calculated as an integral over the three-body phase space from
the product of the $X$ wave function and the decay vertex $D^{*0}\to
D^0\pi^0$. The result of Ref.~\cite{Vol2004} is improved in Ref.~\cite{pions}
by taking next-to-leading (NLO) corrections.  The counting scheme used in
Ref.~\cite{pions} treats the pion effects perturbatively in distinction from
the nonperturbative inclusion of the LO contact operators in analogy to what
was proposed for the NN system in Refs.~\cite{KSW}. The results of these two
calculations (taken from Fig.~7 of Ref.~\cite{pions}) are shown in the right
panel of Fig.~\ref{width} by the dot-dashed curve (LO) and by the blue band
(NLO), respectively.  
For the sake of
comparison our result\ftnote{1}{We would like to stress that in both above-mentioned
calculations the $X$ is assumed to be a genuine bound (not a resonance)
state. In the full treatment, however, such a bound state does not exist, so
that one cannot calculate the width directly, following the lines of
Refs.~\cite{Vol2004,pions}. Instead, one is left to extract the $X$ width
using the procedure described above in this chapter.} is given in the right panel of Fig.~\ref{width} as the
(red) solid and dashed lines, where the former is the result for the full
calculation, while the latter is the result when the  $D\bar
D\gamma$ channel is omitted---since this contribution to the
width was not included in the $X$-EFT calculations, it is this curve that is
to be compared to the results of Refs.~\cite{Vol2004,pions}.
We are therefore to conclude that the $X$-EFT reproduces the results of
our more complete calculation very nicely. 
This level of agreement provides strong support of the power counting
underlying the $X$-EFT. 

All the results shown in Figs.~\ref{rate} and \ref{width}  were calculated with $\Lambda=500$ MeV, and the value of
$C_0$ for the full problem (case 3) was chosen to be negative, see solid (red) branch in the right panel of Fig.~\ref{c0fig}.
Note, however, that
the variation of the results with the cutoff is very mild for the range of $\Lambda$ from 300 to 1700~MeV and from
2500 to 3800~MeV.
For example,
varying $\Lambda$ in the range $300~\mbox{MeV}\lesssim\Lambda\lesssim 1700$~MeV influences the width only at a few percent level
comparable with the numerical noise. Also we obtain that choosing another branch of $C_0$ results in the uncertainty
in the width of a similar (small) size.

As the final remark let us mention that, aiming at the effects related to dynamical pions, we considered only the part of the $X(3872)$ width that comes from the $D\bar{D}\pi$ intermediate state. The physical $X(3872)$ supposedly acquires a
significant part of its width from inelastic channels like $J/\psi \pi\pi$ and $J/\psi \pi\pi\pi$.
However, the effects discussed here should be of relevance for the line shapes
of the $X(3872)$ in the $D\bar{D}\pi$ channel.

\section{Summary}

We investigated the role played by dynamical pions on the structure of the
$X(3872)$ by solving the full three-body $D\bar D\pi$ problem. The two-body $\pi D\to \pi D$ input is fixed by the width of the $D^*$, however,
a short-ranged $D\bar{D}^*\to D\bar{D}^*$ contact term needed to be introduced to
arrive at well defined equations. Depending on the regulator used
for solving the equations, the contact term can take values between
$-\infty$ and $+\infty$ for the cut off regularisation
scheme. On the other hand, the strength of the one-pion
exchange is fixed. Thus, the relative importance of the pion exchange as compared
to the short-ranged contributions to the $D \bar D^*$ scattering potential appears to be strongly
regulator-dependent. We are therefore led to conclude that no model independent statement on the
importance of the one-pion exchange for the formation of the $X(3872)$ is possible, contrary to
earlier claims~\cite{tornqvist2,ThCl}.

In addition, we found that the residue for $X\to D\bar{D}^*$ is weakly dependent
on the kind of pion dynamics included. Especially, the dependence
of the residue on the $X$ binding energy is very close for a fully dynamical
calculation and for a calculation with a contact-type interaction only.
A deviation between the coupled-channel and the single-channel treatment is clearly observed
but with the larger effect for binding energies beyond 1~MeV.

The most striking effect of dynamical pions is observed in their impact
on the $X$ line shapes:
in the fully dynamical calculation the width from the $D\bar{D}\pi$ intermediate
states appears to be reduced by about a factor of 2, from 102 down to 44 keV, assuming that the
$X(3872)$ corresponds to a resonance state with a peak at 0.5 MeV below the $D \bar D^*$ threshold.
Stated differently, by using the naive static approximation for the $D\bar{D}\pi$ intermediate states one overestimates substantially their effect on the $X$ width.
Although the total width of the $X$ is rather saturated by inelastic channels like $J/\psi \pi\pi$ and $J/\psi
\pi\pi\pi$, our findings should be of relevance for the predicted line shape in the
$D^0\bar{D}^0\pi^0$ channel, where the signal below the $D\bar{D}^*$ threshold
is controlled exactly by the $D\bar{D}\pi$ cuts. For example, if the line shape near threshold is analysed
within the scattering length approximation, the imaginary part of the inversed scattering
length, $\gamma_2$, which is the only parameter affected by the three-body dynamics, has to be taken from
the fully dynamical calculation. The parameter $\gamma_2$ (and thus the role
of pion dynamics) appears to be a smooth but vivid function of the binding energy, for which we provide a simple 
analytic parameterisation suitable to mimic three-body effects numerically in the data analysis.

On the contrary, the effect of the coupled-channel dynamics on the $X$ width turned out to be
rather moderate, which can be attributed to the fact that both the real part of the resonance 
pole $E_B$ and the $X$ width $\varGamma_X$ are small as compared to the separation $\Delta M$ 
between the neutral and the charged thresholds.

Our results for the $X$ width and for the residue appear to be in good agreement with those obtained in the $X$-EFT
approach~\cite{pions,Vol2004}, which justifies the central assumption underlying the $X$-EFT that
pions can be treated perturbatively. In addition, an important progress achieved in our work due 
to considering the full coupled-channel dynamics with nonperturbative pions is that
pion range corrections are included to all orders in our calculation. Because of the proximity of the
$D\bar{D}\pi$ threshold, those are expected to give a prominent contribution to
the range corrections even beyond leading order. 
We therefore expect the nonperturbative calculation to have a smaller uncertainty than that with the perturbative
treatment of pions. In addition, the $X$ pole is located very close to both the $D\bar{D}\pi$ as
well as the $D\bar{D}^*$ thresholds, which all influence the pertinent integrals. 
Thus, although for the $X$ the effects can be nicely absorbed into a counter
term, this is not necessarily the case anymore for other, related resonances,
say in the $B$ sector\ftnote{2}{A first step towards a combined study of $D^{(*)}\bar D^{(*)}$ and  $B^{(*)}\bar
B^{(*)}$ was taken in Ref.~\cite{manolo}.}. In this
sense our calculation appears as an important step for a common understanding
of a larger class of quarkonium resonances.

\begin{acknowledgments}
The authors would like to thank E.~Braaten for reading the manuscript and for valuable comments. 
The work was supported in parts by funds provided from the Helmholtz
Association (via Grants No. VH-NG-222 and No. VH-VI-231), by the DFG (via Grants No. SFB/TR 16
and No. 436 RUS 113/991/0-1), by the EU HadronPhysics2 project, by the RFFI (via Grants No. RFFI-09-02-91342-NNIOa and
No. RFFI-09-02-00629a), and by the State Corporation of Russian Federation ``Rosatom.''
\end{acknowledgments}

\appendix

\section{Details of the derivation of the three--body equation (\ref{a})}\label{apt}

After excluding the third equation, the systems (\ref{s10}) and (\ref{s20}), with all arguments and sub(super)scripts restored, read
\be
\left\{
\begin{array}{rcl}
\ds t_{22}^{mn}(\vep,\vep',E)&=&\ds
-\Sigma(p)\delta^{mn}\delta(\vep-\vep')+
\frac{\Sigma(p)}{D_2(\vep)}t_{22}^{mn}(\vep,\vep',E)\\[2mm]
&+&\ds g^2\int d^3q\frac{q_m (\alpha q_p +\beta p_p)}{D_3(\vep,\veq)
D_2(-\veq -\alpha \vep)}t_{\bar 2 2}^{pn}(-\veq -\alpha
\vep,\vep',E)\\[7mm]
\ds t_{\bar 2 2}^{mn}(\vep,\vep',E)&=&\ds -g^2\frac{(\alpha p_m+p'_m)(\alpha
p'_n+p_n)}{D_3(\vep,-\alpha \vep -\vep')}
+\frac{\Sigma(p)}{D_2(\vep)}t_{\bar 2 2}^{mn}(\vep,\vep',E)\\[2mm]
&+&\ds g^2\int d^3q\frac{q_m(\alpha q_p +\beta p_p)}{D_3(\vep,\veq)
D_2(-\veq -\alpha \vep)}t_{2 2}^{pn}(-\veq -\alpha \vep,\vep',E),
\end{array}
\right.
\label{s1}
\ee
\be
\left\{
\begin{array}{rcl}
\ds t_{\bar 2 \bar 2}^{mn}(\bar {\vep},\bar {\vep}',E)&=&\ds
-\Sigma(\bar{p})\delta^{mn}\delta(\bar {\vep}-\bar {\vep}')+
\frac{\Sigma(\bar{p})}{D_2(\bar
{\vep})}t_{\bar 2 \bar 2}^{mn}(\bar {\vep},\bar {\vep}',E)\\[2mm]
&+&\ds g^2\int d^3\bar {q}\frac{\bar{q}_m (\alpha \bar{q}_p+\beta
\bar{p}_p)}{D_3(\bar {\vep},\bar {\veq})
D_2(-\bar {\veq} -\alpha \bar {\vep})}t_{2 \bar 2}^{pn}(-\bar
{\veq}-\alpha
\bar {\vep},\bar {\vep}',E)\\[7mm]
\ds t_{2 \bar 2}^{mn}(\bar {\vep},\bar {\vep}',E)&=&\ds
-g^2\frac{(\alpha\bar{p}_m+\bar{p}'_m)(\alpha\bar{p}'_n+\bar{p}_n)}{D_3(\bar {\vep},-\alpha \bar {\vep}
-\bar {\vep}')}+\frac{\Sigma(\bar{p})}{D_2(\bar {\vep})}t_{2
\bar 2}^{mn}(\bar {\vep},\bar {\vep}',E)\\[2mm]
&+&\ds g^2\int d^3\bar {q}\frac{\bar{q}_m(\alpha\bar{q}_p+\beta\bar{p}_p)}{D_3(\bar {\vep},\bar {\veq})
D_2(-\bar {\veq} -\alpha \bar {\vep})}t_{\bar 2 \bar 2}^{pn}(-\bar
{\veq} -\alpha \bar {\vep},\bar {\vep}',E).
\end{array}
\right.
\label{s2}
\ee

It can be demonstrated then that $t_{23}^{m}(\vep;\vep',\veq')$ and $t_{\bar 2 3}^{m}(\bar {\vep};\bar
{\vep}',\bar {\veq}')$ of the form
\bea
t_{23}^{m}(\vep;\vep',\veq';E)&=&gq'_m\delta(\vep-\vep')-\frac{g}{D_2(\vep')}t_{22}^{mn}(\vep,\vep',E)q'_n
\nonumber\\
&-&\frac{g}{D_2(-\veq'-\alpha \vep')}t_{\bar 2 2}^{mn}(\vep,-\veq'-\alpha \vep',E)(\alpha q'_n+\beta p'_n),
\label{t23}\\
t_{\bar 2 3}^{m}(\bar {\vep};\bar {\vep}',\bar {\veq}';E)&=&g\bar{q}'_m\delta(\bar {\vep}-\bar {\vep}')
-\frac{g}{D_2(\bar {\vep}')}t_{22}^{mn}(\bar {\vep},\bar {\vep}',E)\bar{q}'_n\nonumber\\
&-&\frac{g}{D_2(-\bar {\veq}'-\alpha \bar {\vep}')}t_{\bar 2 2}^{mn}(\bar {\vep},-\bar {\veq}'-\alpha \bar {\vep}',E)
(\alpha \bar{q}'_n+\beta\bar{p}'_n).
\label{tbar23}
\eea
satisfy the last system, given by Eq.~(\ref{s30}).

Equations for the $C$-even and $C$-odd $D \bar D^*$ matrix elements
\be
t_{\pm}=t_{22}\pm t_{\bar{2}2},
\ee
take the form:
$$
t_{\pm}^{mn}(\vep,\vep',E)=-\frac{\Sigma(p)D_2(\vep)}{\Delta(p)}
\delta^{mn}\delta(\vep-\vep') \mp g^2\frac{(\alpha p_m+p'_m)(\alpha p'_n+p_n)}{D_3(\vep, -\alpha \vep
-\vep')}\frac{D_2(\vep)}{\Delta(p)}
$$
\be
\pm g^2\frac{D_2(\vep)}{\Delta(p)}\int d^3s\frac{(s_m+\alpha p_m)(\alpha s_p+p_p)}{D_3(\vep,-\ves
-\alpha \vep)D_2(\ves)}
t_{\pm}^{pn}(\ves,\vep',E),
\ee
which can be brought to the form of Eq.~(\ref{a}) by the substitution
\be
t_{\pm}^{mn}(\vep,\vep',E)=-\frac{\Sigma(p)D_2(\vep)}{\Delta(p)}
\delta^{mn}\delta(\vep-\vep')+
\frac{D_2(\vep)}{\Delta(p)}a_{\pm}^{mn}(\vep,\vep',E)
\frac{D_2(\vep')}{\Delta(p')}.
\label{intra}
\ee

\end{document}